\documentclass[useAMS,usenatbib]{mn2e}

\newcommand{\beq}{\begin{equation}}
\newcommand{\eeq}{\end{equation}}
\newcommand{\beqa}{\begin{eqnarray}}
\newcommand{\eeqa}{\end{eqnarray}}
\newcommand{\so}{_{\cal O}}

\usepackage{graphicx}
\usepackage{type1cm}
\usepackage{eso-pic}
\usepackage{color}
\usepackage{amsmath}

\title[Redshift and distances in inhomogeneous $\Lambda$CDM]{Redshift and distances in a $\Lambda$CDM cosmology with non-linear inhomogeneities}
\author[N. Meures and M. Bruni]{Nikolai Meures\thanks{E-mail:
nikolai.meures@port.ac.uk} and Marco Bruni\thanks{E-mail: marco.bruni@port.ac.uk}\\
Institute of Cosmology and Gravitation, University of Portsmouth, Portsmouth PO1 3FX, UK}

\date{\today}

\pagerange{\pageref{firstpage}--\pageref{lastpage}} \pubyear{2011}
\begin{document}

\label{firstpage}
\maketitle
\begin{abstract}
Motivated by the dawn of precision cosmology and the wealth of forthcoming high precision and volume galaxy surveys, in this paper we study the effects of inhomogeneities on light propagation in a flat $\Lambda$CDM background. To this end we use exact solutions of Einstein's equations \citep{MeuBru11-1} where, starting from small fluctuations, inhomogeneities arise from a standard growing mode and become non-linear. While the matter distribution in these models is necessarily idealised, there is still enough freedom to assume an arbitrary initial density profile along the line of sight. We can therefore model over-densities and voids of various sizes and distributions, e.g. single harmonic sinusoidal modes, coupled modes, and more general distributions in a $\Lambda$CDM background. Our models allow for an exact treatment of the light propagation problem, so that the results are unaffected by approximations and unambiguous.  Along lines of sight with density inhomogeneities which average out on scales less than the Hubble radius, we find the distance redshift relation to diverge negligibly from the Friedmann-Lemaitre-Robertson-Walker (FLRW) result. On the contrary, if we observe along lines of sight which do not have the same average density as the background, we find large deviations from the FLRW distance redshift relation. Hence, a possibly large systematic might be introduced into the analysis of cosmological observations, e.g.\ supernovae, if we observe along lines of sight which are typically more or less dense than the average density of the Universe. In turn, this could lead to wrong parameter estimation: even if the Cosmological Principle is valid, the identification of the true FLRW background in an inhomogeneous universe maybe more difficult than usually assumed.
\end{abstract}


\section{Introduction}
At the very basis of modern cosmology lies the assumption the Universe is, at any given time, homogeneous and isotropic on large scales. This is translated mathematically into a Robertson-Walker metric, i.e. a metric that is assumed to represent a space average and is therefore exactly homogeneous and isotropic. In addition, the non-trivial hypothesis is made that this metric should be a solution of Einstein's equations, thereby giving rise to a FLRW universe model. Having assumed the Cosmological Principle, the growth of inhomogeneities and their effects are typically modelled with perturbation theory about a ``background'' FLRW model. Within this framework, the formation of non-linear structures at smaller scales is considered in N-body simulations using the Newtonian approximation \citep{Spr05}. Most observations are interpreted assuming this Friedmannian framework; in particular, distances are computed assuming a FLRW distance-redshift relation, i.e. completely neglecting inhomogeneities. 

We are currently living in a time when galaxy surveys and other observations are reaching unprecedented sky coverage and precision; therefore, it seems timely to fully investigate the effects of the non-linear growth of structures on observations, within a general relativistic framework. Much work has been done in trying to understand the effect of inhomogeneities on observations by using perturbations around an FLRW model \citep{DyeRoe72, DyeRoe73, DyeRoe74, Sas87, FutSas89, PynBir04, BarMatRio05, BonDurGas06, Ras11-2}. However, these analyses are limited to linear structure growth and therefore cannot properly take into account non-linear inhomogeneities. Non-linearities can be modelled using idealised matter distributions. Several different approaches using generalised Swiss-Cheese models or alternative geometries have been considered in e.g.\ \citet{BroTetTza07, Mar07, BroTetTza08, BisNot08, CliZun09, BolCel10, Szy11, NwaIshTho11}, where significant deviations are usually only found for very large scale density inhomogeneities. On the contrary, claims are made that small scale structure formation might have a back-reaction effect on the overall expansion of the Universe, see \citet{Ras06} and \citet{Buc08} for overviews of this topic. We do not investigate this phenomenon of back-reaction in this paper though, as our model clearly splits into inhomogeneities and background dynamics.

A class of models which are very appropriate for considering a discrete distribution of matter in an otherwise FLRW expanding universe has recently been analysed in depth by \citet{CliFer09, Clifer09-2}. In these models, originally introduced by \citet{LinWhe57} and revised by \citet{Red88}, matter is described by pointlike masses in a spherically symmetric void box (represented by Schwazschild space-times) and these boxes are distributed in a lattice and the overall expansion is described by the Friedmann equation. The main motivation of Clifton and Ferreira in following the Lindquist and Wheeler construction is the observation that the Universe largely consists of galaxies and clusters of galaxies surrounded by vacuum. The question they address is how observations and measurements of the cosmological parameters are affected in a highly inhomogeneous universe whose overall dynamics are homogeneous and isotropic. However, this lattice construction is only an approximate solution to Einstein's equations and has regions of ``no man's land" in between the matched spheres which might have an effect on the light tracing, see \citet{Cli10}. Perhaps even more importantly, the inhomogeneities in this model are strongly non-linear at all times. 

Having the same type of questions addressed by \citet{CliFer09} in mind, in this paper we shall investigate the optical properties of an exact solution to Einstein's field equations (EFEs), developed in our previous paper \citep{MeuBru11-1}. This solution belongs to the second class Szekeres models \citep{Sze75} for dust, generalised by \citet{BarSte84} to include the cosmological constant $\Lambda$. In \citet{MeuBru11-1} we used the \citet{GooWai82} formulation of the Szekeres models to reconsider the Barrow and Stein-Schabes solutions in a $\Lambda$CDM context. In our model, starting from standard small perturbations of a $\Lambda$CDM homogeneous and isotropic universe, the matter distribution is continuous and can evolve to a highly non-linear stage. In the process, the inhomogeneities can either form a distribution of large voids, over-densities, or a mixture of the two with over-densities possibly even forming pancakes as in the Zel'dovich approximation in Newtonian cosmology. The benefit of our model is therefore two-fold: \textit{i}) we consider exact solutions of Einstein's equations, therefore avoiding any possible problem associated with approximations and matching and \textit{ii}) these exact solutions describe non-linear inhomogeneities growing on top of a FLRW background with the possibility of modelling a rather arbitrary distribution of both voids and over-densities.

In particular, we shall consider the effect of inhomogeneities on the redshift, angular diameter distance and distance modulus. We begin by considering single mode harmonic sinusoidal deviations from homogeneity and then the case of coupled modes. Finally, closer in spirit to the \citet{CliFer09} work, we consider inhomogeneities where peaks and voids, arranged in a periodic array, are more strongly separated than can be achieved by simple harmonic distributions. We demonstrate that the deviations from the FLRW background in the determination of the distances is mainly due to the Ricci and Weyl focusing terms in the Sachs equations and show that instead the shear of the null congruence has a negligible effect. We also briefly investigate the effect of mode coupling on the growth of structure and, interestingly, we show that even a long wavelength mode with small amplitude can strongly enhance the growth of short wavelength modes, thanks to the non-linear coupling. The non-linear interaction of different modes does not seem to influence the distance measures significantly and we find that changes in the redshift and distances are mostly affected by the long wavelength modes. Using an array of density profiles which are not sinusoidal but quite peaked around the maximum and separated by large voids, we find that the effect on the redshift and distance measures does not prove to be significantly more than using an initially sinusoidal density distribution with the same wavelength of the array scale. Overall, we find that all deviations in the redshift and distance measures are less than 1\%, when we consider what we refer to as ``compensated inhomogeneities" along the line of sight, i.e.\ where the average density along the line of sight matches the background density. However, this does not need to be the case: when the inhomogeneities are on average above or below the background, the effects on redshift and distance measures can be very large. 

A summary of the paper is as follows. In Sec.\ \ref{sec:sol} we will briefly present the exact solution we will be using in this work, referring the reader to \citet{MeuBru11-1} for more details. Subsequently, in Sec.\ \ref{sec:geo}, we will derive the null geodesic equations for the given metric and in Sec.\ \ref{sec:opt} we derive the form that the Sachs optical equation will take for a given physical situation that is to be investigated. In Sec.\ \ref{sec:res} we shall present the results of our analysis, considering single mode deviations in Sec.\ \ref{sec:res1}; multiple modes and their coupling and effects in Sec.\ \ref{sec:res2}; and the effects of an array of strong peaks and large voids in Sec.\ \ref{sec:res3}. In Sec.\ \ref{sec:conc} we draw our conclusions. In Appendix \ref{sec:trans} we present details on the tetrad transformations needed to derive the Sachs optical equations in our model.

Throughout the paper we choose units $c=8\pi G=1$ and assume the standard $\Omega_{\Lambda}=0.75$ and in the commonly used units $H_0=72\text{km}\text{s}^{-1}\text{Mpc}^{-1}$.

\section{An exact solution of sufficient generality}\label{sec:sol}
In this section, we would like to present a short summary of the main results of our previous paper \citep{MeuBru11-1}, where we developed a class of exact solutions to EFEs,
\beq\label{eq:EFEs}
 G_{ab}=T_{ab}-\Lambda g_{ab},
\eeq
for the metric
\beq\label{eq:line}
 ds^2=-dt^2+S(t)^2\left[dx^2+dy^2 + Z(\textbf{x},t)^2dz^2\right],
\eeq
where we are using synchronous comoving coordinates, so that $t$ represents a universal cosmic time. The variable $\textbf{x}$ denotes all three spatial Cartesian coordinates $x$, $y$ and $z$. Note that whenever we will use the variable $z$, we will be referring to the coordinate; whereas we will refer to the redshift as $z^{IN}$ or $z^{FLRW}$, respectively for an inhomogeneous or FLRW universe. We solve Eq.\ (\ref{eq:EFEs}) for an irrotational pressureless fluid with four velocity $u^a=\delta^a_0$, for which 
\beq\label{eq:energy}
 T^{00}=\rho,
\eeq
and all other components of $T^{\mu\nu}$ are zero. The solutions for a metric of this type for pure dust were first introduced by \citet{Sze75} and then brought into a similar notation that we are using here and analysed by \citet{GooWai82,GooWai82-1}. The solution for dust and a cosmological constant was found by \citet{BarSte84}. We are here considering one specific sub-class of these models, usually called the second class Szekeres models, and we choose the FLRW background to be spatially flat. For detailed accounts of exact solutions in general relativity, see \citet{Kra97}, \citet{Ste03} and \citet{BolKraHelCel09}.

The function $S(t)$ is the scale factor of a FLRW $\Lambda$CDM background, with Friedmann equation
\beq\label{eq:Friedmann}
 \frac{\dot{S}^2}{S^2}=\frac{\bar{\rho}_0}{3S^3}+\frac{\Lambda}{3},
\eeq
which admits the solution
\beq\label{eq:solS}
 S(t)=\left( \frac{1-\Omega_{\Lambda}}{\Omega_{\Lambda}}\right)^{1/3}\sinh ^{2/3} \left(\frac{3}{2}H_0\sqrt{\Omega_{\Lambda}}t\right).
\eeq
Here $\bar{\rho}=\bar{\rho}_0/S^3$ is the background energy-density and we have normalised $S=1$ today. We have used the standard parametrisation $\Omega_m=\bar{\rho}_0/(3H_0^2)$ and $\Omega_{\Lambda}=\Lambda/(3H_0^2)$, where $H_0$ is the Hubble parameter and $\Omega_m=1-\Omega_{\Lambda}$.\\

Note that we choose $S$ as well as the function $Z$ in (\ref{eq:line}) to be dimensionless, so that the coordinates have a dimension of length.

The function $Z$ in the line element (\ref{eq:line}) can be split as
\beq\label{eq:decoZ}
Z(\textbf{x},t)=F(z,t)+A(\textbf{x}),
\eeq
where $A$ can be written in the form
\beq\label{eq:decoA}
A(\textbf{x})=1+B\beta_+(z)\left\{ \left[x+\gamma(z)\right]^2+\left[y+\omega(z)\right]^2\right\},
\eeq 
where 
\beq\label{eq:B}
B=\frac{3}{4}H_0^2\left[\Omega_{\Lambda}(1-\Omega_{\Lambda})^2\right]^{1/3},
\eeq
with the free functions $\beta_+(z)$, $\gamma(z)$ and $\omega(z)$.

\begin{figure}
\begin{center}
$\begin{array}{ll}
\includegraphics[width=0.4\textwidth]{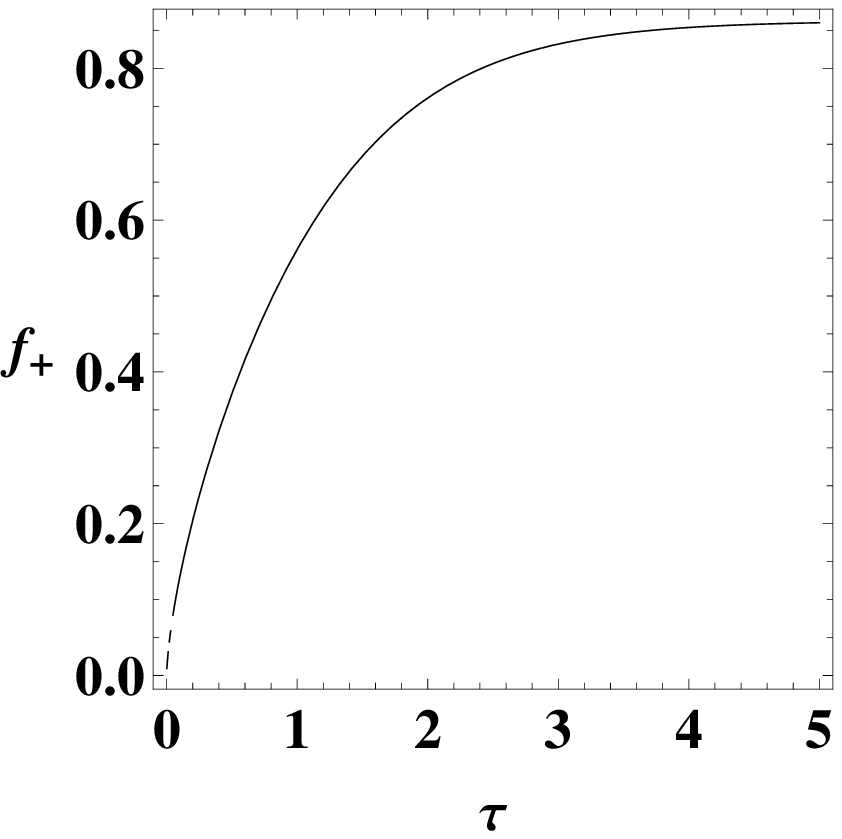} \\
\includegraphics[width=0.4\textwidth]{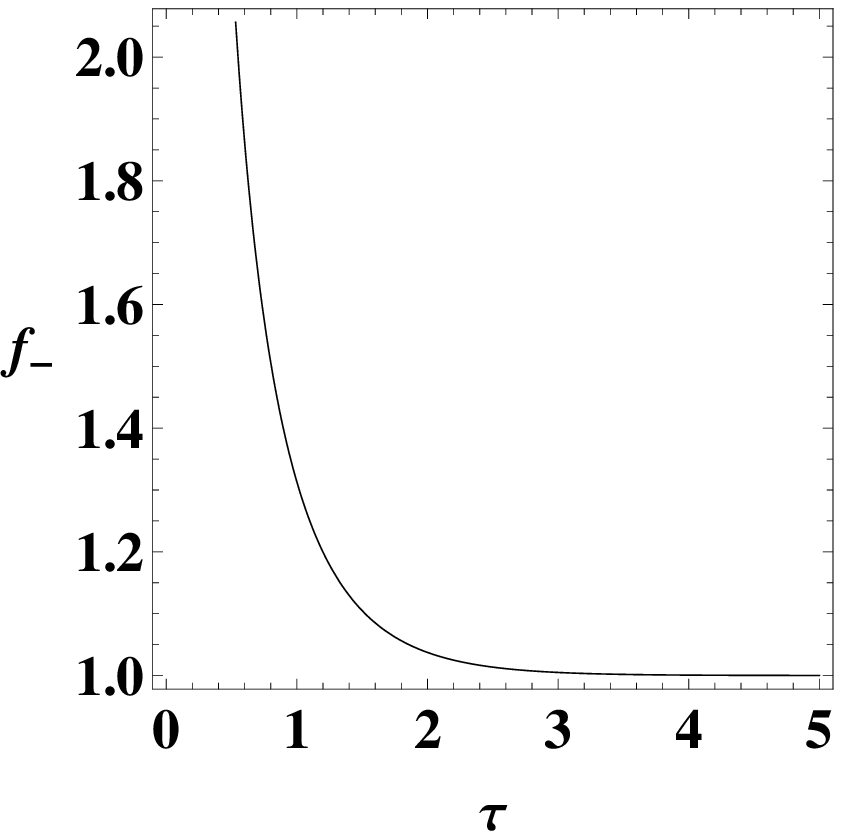}  \\
\end{array}$
\end{center}
\caption{Plots of the growing (top panel) and decaying (bottom panel) modes of the solution for $F$, as derived from Eq.\ (\ref{eq:diffFhom}). The solutions plotted here are given in Eqs.\ (\ref{eq:solF-}) and (\ref{eq:solF+}).}
\label{fig:growdecay}
\end{figure}

Remarkably, $F$ obeys the second order linear homogeneous ordinary differential equation
\beq\label{eq:diffFhom}
 \ddot{F}+2\frac{\dot{S}}{S}\dot{F}-\frac{\bar{\rho}}{2}F=0,
\eeq
which is exactly the same equation that $\delta=(\rho-\bar{\rho})/\bar{\rho}$ satisfies in Newtonian linear perturbation theory, as well as in the synchronous comoving gauge for relativistic perturbations of FLRW. This equation admits two linearly independent solutions and hence we write
\beq\label{eq:decoF}
 F(z,t)=\beta_+(z)f_+(t)+\beta_-(z)f_-(t),
\eeq
where $\beta_+$ and $\beta_-$ are free functions of the coordinate $z$ and $f_+$ is the growing mode and $f_-$ is the decaying mode of the solution, which we find to be
\begin{subequations}
\beqa
 f_-&=&\frac{\cosh(\tau)}{\sinh(\tau)},\label{eq:solF-}\\ 
f_+ &=&\frac{\cosh(\tau)}{\sinh(\tau)}\int\frac{\sinh^{2/3}(\tau)}{\cosh^2(\tau)}d\tau\label{eq:solF+},
\eeqa
\end{subequations}
where we defined the dimensionless variable $\tau=\frac{3}{2}H_0 \sqrt{\Omega_{\Lambda}}t$. The two independent solutions for $F$ are shown in Fig.\ (\ref{fig:growdecay}). In the matter dominated era, when the effect of $\Lambda$ is negligible, $f_+(t)\propto S(t)\propto t^{2/3}$ and $f_-\propto t^{-1}$, as it is well known, see e.g.\ \citet{Pee80}.

With all the free functions having been identified and having solved for all the time dependent functions, we can find the expressions for the density $\rho$, the background density $\bar{\rho}$ and density deviation $\delta$ to be
\beqa
 \rho& =& \frac{\bar{\rho}_0A}{S^3(F+A)},\label{eq:rhomaster}\\
\bar{\rho}&=&\frac{\bar{\rho}_0}{S^3},
\eeqa
and 
\beqa
 \delta&\equiv&\frac{\rho-\bar{\rho}}{\bar{\rho}},\\
 &=&-\frac{F}{Z}=-\frac{F}{F+A}.
 \eeqa
As we are only concerned with late times here, we will not be considering the decaying mode of $F$, $\beta_-f_-$, as it would not effect our results, but complicate the calculations. This can easily be done by choosing $\beta_-=0$ in (\ref{eq:decoF}). On the other hand, a very useful feature of our solution is that $F$ satisfies the linear differential equation (\ref{eq:diffFhom}), so that a superposition principle applies, and that the arbitrariness of the growing mode function $\beta_+$ in (\ref{eq:decoF}) allows us to construct an arbitrary matter distribution along $z$. We find that at early times, along the $z$-axis, $\delta\approx -F$ and so choosing a function $\beta_+$ directly determines the initial matter distribution along the $z$-axis. Choosing $\beta_+=A\sin(kz)$ and therefore $\delta\propto\sin(kz)$ initially for some amplitude $A$ and some wavelength $k$, implies that $\beta_+$ and hence the initial distribution of $\delta$ are periodic on a scale of $2\pi/k$. This situation is what we will be referring to as a compensated density deviation later on in the paper, as averaging the density deviation $\delta$ along the $z$-axis at early times, would tend to a zero average once the period of the deviations is reached. In comparison, we will be referring to over-densities for $\beta_+<0$, i.e.\ $\delta>0$ and under-densities for $\beta_+>0$ and hence $\delta<0$.

\begin{figure}
\includegraphics[width=0.45\textwidth]{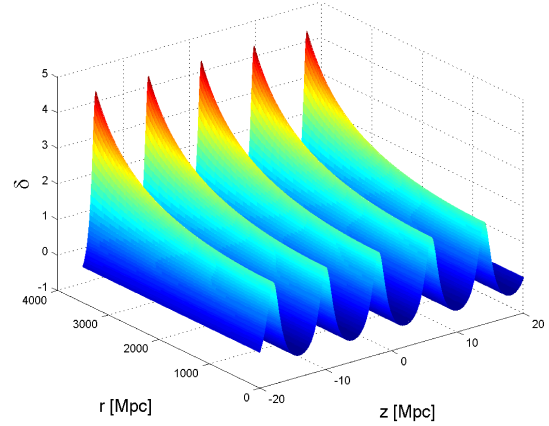}
\caption{Density deviation profile today, corresponding to an initial density perturbation of $\delta\propto \sin(kz)$ for $k=2\pi/8\mathrm{Mpc}^{-1}$ along the $z$-axis, for $\gamma=\omega=0$. We only indicate the coordinate distance $r=\sqrt{x^2+y^2}$ from the $z$-axis, as the solution for $\gamma=\omega=0$ is symmetric about this axis.}
\label{fig:3d}
\end{figure}

\begin{figure}
\includegraphics[width=0.45\textwidth]{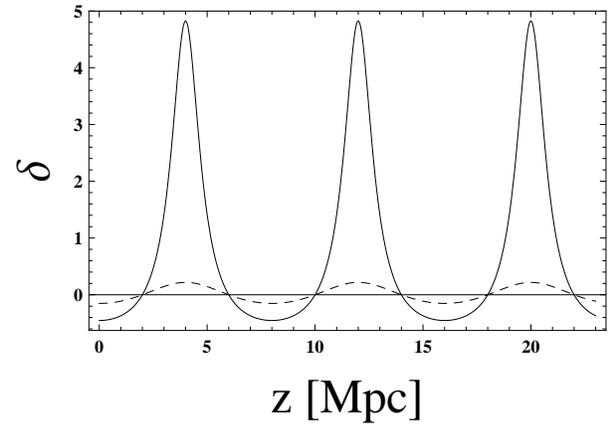}
\caption{Density deviation profile along the $z$-axis at different times for an initial density perturbation $\delta\propto \sin(kz)$ for $k=2\pi/8\mathrm{Mpc}^{-1}$ along the $z$-axis, for $\gamma=\omega=0$. The solid line shows the profile of the inhomogeneities today and the dashed line shows the same inhomogeneities at a redshift $z^{FLRW}=5$. This shows how an initial sinusoidal deviation in the density changes its profile at late times due to the non-linear growth of the inhomogeneities.}
\label{fig:delta}
\end{figure}

To aid the reader in gaining an intuitive understanding of the density profiles that we will be using in this paper, we have included two figures with density profiles for $\gamma=\omega=0$. In Fig.\ \ref{fig:3d} we show what the shape of the density deviation would be today, if we chose the initial profile to be $\delta\propto \sin(kz)$ for $k=2\pi/8 \mathrm{Mpc}^{-1}$ along the $z$-axis. Only two spatial dimensions are displayed here, because of the symmetry of the model around the $z$-axis for $\gamma=\omega=0$. Note that in Figs.\ \ref{fig:3d}, \ref{fig:delta} and \ref{fig:deltas} we use coordinate distances, measured in Mpc, which serve as a reference distance. Given our metric (\ref{eq:line}) and our choice $S=1$ today, the coordinate distance in directions orthogonal to the $z$-axis equals the proper distance today, but in any other direction the proper distance will involve an integration over the function $Z$. 

In Fig.\ \ref{fig:delta} we display how an initial sinusoidal density deviation grows non-linearly into a shape with voids and high over-density peaks. While the metric function $F$, initially $F\propto\delta$, evolves linearly and remains sinusoidal, $\delta$ can grow strongly, even developing pancakes, see \citet{MeuBru11-1}.

\section{The null geodesic equations}\label{sec:geo}
We now want to calculate the null geodesic equations using the Euler-Lagrange formalism. We start from the Lagrangian
\beq
L= -\left(\frac{d t}{d\lambda}\right)^2 +S^2\left\{\left[\left(\frac{d x}{d\lambda}\right)^2+\left(\frac{d y}{d\lambda}\right)^2\right]+Z^2\left(\frac{d z}{d\lambda}\right)^2 \right\},
\eeq
where $\lambda$ is an affine parameter. From this we find the first null geodesic equation to be
\beqa
S\dot{S}&\left[\left(\frac{d x}{d\lambda}\right)^2 +\left(\frac{d y}{d\lambda}\right)^2+Z^2\left(\frac{d z}{d\lambda}\right)^2\right]\nonumber\\
&+Z\beta_+\dot{f}_+S^2\left(\frac{d z}{d\lambda}\right)^2=-\frac{d^2t}{d\lambda ^2},
\eeqa
where an over-dot denotes differentiation with respect to $t$. The second and third equations take the similar form

\beq
\frac{d ^2x}{d\lambda^2}+2\frac{\frac{d S}{d\lambda}}{S}\frac{d x}{d\lambda}-2BZ\beta_+(x+\gamma)\left(\frac{d z}{d\lambda}\right)^2=0,
\eeq
and
\beq
\frac{d ^2y}{d\lambda^2}+2\frac{\frac{d S}{d\lambda}}{S}\frac{d y}{d\lambda}-2BZ\beta_+(y+\omega)\left(\frac{d z}{d\lambda}\right)^2=0,
\eeq
the last null geodesic equation, governing the photon's motion along the $z$-axis, is given by
\beqa
\frac{d ^2z}{d\lambda^2}&=&-\frac{\left(\frac{d z}{d\lambda}\right)^2}{Z}( (\beta_+)_z\left\{f_++B\left[(x+\gamma)^2+(y+\omega)^2\right] \right\}\nonumber\\
 &+&2B\beta_+\left[(x+\gamma)\gamma_z+(y+\omega)\omega_z \right] )\nonumber\\
&-&2\frac{d z}{d\lambda}\left[\frac{\frac{d S}{d\lambda}}{S}\right.\nonumber\\
&+&\left.\beta_+\frac{\frac{d f_+}{d\lambda}+2B(x+\gamma)\frac{d x}{d\lambda}+2B(y+\omega)\frac{d y}{d\lambda}}{Z} \right],
\eeqa
where a subscript $z$ denotes differentiation with respect to the Cartesian coordinate $z$. These differential equations contain derivatives with respect to three different variables, $t$, $\lambda$ and $z$. Since only the free functions are functions of $z$ and we will specify them on a case by case basis, we can consider these derivatives to be known. We would like to consider functions of only one variables and not two, $\lambda$ and $t$. Since we are considering null geodesics here, we can specify $\lambda$, such that 
\beq
\frac{d}{d\lambda}=E\frac{d}{dt},
\eeq
where $E=-u^al_a$ is the energy of the photon. Using this relationship and a new time variable 
$\tau=\frac{3}{2}H_0 \sqrt{\Omega_{\Lambda}}t$ we can simplify the set of differential equations to
\beqa
-\frac{E'}{E}&=&S^2\left(\frac{9}{4}H_0^2\Omega_{\Lambda}\right)\{\nonumber\\
&&z'^2Z\left[Z\frac{S'}{S}+f'_{+}\beta_+ \right] +\frac{S'}{S}(x'^2+y'^2)\},\label{eq:odeE}
\eeqa

\beq\label{eq:odex}
x''+\left(2\frac{S'}{S}+\frac{E'}{E}\right)x'-2BZ\beta_+(x+\gamma)z'^2=0,
\eeq

\beq\label{eq:odey}
y''+\left(2\frac{S'}{S}+\frac{E'}{E}\right)y'-2BZ\beta_+(y+\omega)z'^2=0,
\eeq

\beqa
z''&+&z'^2\frac{Z_z}{Z}+2z'\left[\frac{S'}{S}+\frac{1}{2}\frac{E'}{E}\right.\nonumber\\
&+&\left.\beta_+\frac{f'_++2B(x+\gamma)x'+2B(y+\omega)y'}{Z} \right]=0,\label{eq:oder}
\eeqa
where a dash denotes differentiation with respect to $\tau$. Initial conditions here should be chosen according to the situation that is to be modelled. We will always integrate starting from the observers position (which we denote by ${\cal O}$), which we will therefore place at the origin, $x|\so=y|\so=z|\so=0$ as the position, where the observer is situated, $E|\so=1$ as this is just a normalisation, i.e.\ the redshift in an inhomogeneous universe $z^{IN}=E/E|\so-1$ and thus $E|\so=1$ means that $z^{IN}=E-1$. The initial conditions $x'|\so$, $y'|\so$ and $z'|\so$ are chosen depending on into which spatial direction we would like to perform the light tracing. These null geodesic equations are the most general that we can derive in the given space-time. However, our metric only allows us to freely choose the initial matter distribution along the $z$-axis (for $\gamma=\omega=0$, otherwise along a path dictated by those two functions). Therefore, we will mostly be interested in the propagation of light rays along this `special' $z$-axis. Hence it would be of interest to investigate to what degree the above differential equations simplify, if we only consider the case where $\gamma=\omega=0$ and light rays only travel along the $z$-axis. In this case we find the much reduced system of differential equations

\beq
 -\frac{E'}{E}=\frac{S'}{S}+\frac{F'}{1+F},
 \eeq
 and
 \beq
 z'=\frac{2}{3}\frac{1}{H_0\sqrt{\Omega_{\Lambda}}SZ},
 \eeq
 where we have used the null constraint from the line element to reduce the order of (\ref{eq:oder}). We can use these differential equations to trace single photons into the past, finding their position and energy at any given cosmic time $t$. However, to be able to plot the Hubble diagram, we also need information about how bundles of light rays behave in this space-time and hence we need to consider the Sachs optical equations.

\section{The Sachs optical equations}\label{sec:opt}
To describe the evolution of a bundle of light rays one needs to specify its expansion $\theta$, shear $\sigma$, and rotation $\omega$, which are the quantities whose evolution is described by the Sachs optical equations \citep{Sac61}. However, in this analysis, we closely follow the notation of \citet{Cha92}, who put the optical scalar equations in the context of the Newman-Penrose formalism, see also \citet{Ste03}. Since we are considering point-like sources, we can ignore the rotation $\omega$ of the light bundles. Hence, the Sachs equations take the form
\beqa
\frac{d\theta}{d\lambda}&+&\theta^2+|\sigma|^2=\phi_{00}\label{eq:sachs1}\\
\frac{d\sigma}{d\lambda}&+&2\sigma\theta=\Psi_0,\label{eq:sachs2}
\eeqa
where 
\beq
\phi_{00}=-\frac{1}{2}R_{ab}l^al^b
\eeq
and
\beq
\Psi_0=-C_{abcd}l^am^bl^cm^d.
\eeq
Here $\Psi_0$ and $\phi_{00}$ are, respectively, the zeroth Weyl scalar and one of the Ricci scalars of the Newman-Penrose formalism and represent the Weyl focusing and Ricci focusing in the direction of $l^a$. $R_{ab}$ is the Ricci tensor, $C_{abcd}$ the Weyl tensor, $l^a$ is the affinely parametrised tangent vector to the null geodesic defined as 
\beq
l^a=\frac{dx^a}{d\lambda},
\eeq
and $m^a$ is a complex vector that is orthogonal to $l^a$, null and has magnitude of $1$. The two vectors $l^a$ and $m^a$ are part of a complex Newman-Penrose canonical null tetrad. The expansion $\theta$ and shear $\sigma$ are precisely 
\beqa
\theta&=&\frac{1}{2}l^a_{;a}\\
|\sigma|^2&=&\frac{1}{2}l_{(a;b)}l^{a;b}-\theta^2.
\eeqa
We emphasise that for the moment $l^a$ points in a generic direction and therefore is {\it not} the same $l^a$ as in our previous paper \citep{MeuBru11-1}; however, the advantage of using the canonical Newman-Penrose formalism, as presented in \citet{Cha92}, is that it allows us to easily express $\Psi_0$ and $\phi_{00}$ in the equations above in terms of Weyl and Ricci scalars in the special null tetrad adapted to our metric (see below). The expansion of the bundle of light rays is not a direct observable though and so we would rather like to consider the angular diameter distance $d_A$ and the luminosity distance $d_L$. One finds the relation
\beq
\theta= \frac{\frac{d (d_A)}{d\lambda}}{d_{A}},
\eeq
and Etherington's theorem \citep{Eth33} states that
\beq\label{eq:Eth}
d_L=(1+z^{IN})^2d_A,
\eeq
where $z^{IN}$ is the redshift in a general, inhomogeneous universe. In terms of the angular diameter distance, the Sachs equations take the form
\beqa
\frac{d^2(d_A)}{d\lambda^2}&=&[\phi_{00}-|\sigma|^2]d_A,\\
\frac{d\sigma}{d\lambda}&+&2\frac{\frac{d(d_A)}{d\lambda}}{d_A}\sigma=\Psi_0.
\eeqa
As in Sec.\ \ref{sec:geo}, we will use the time variable $\tau=\frac{3}{2}H_0 \sqrt{\Omega_{\Lambda}}t$ and we also introduce
\beq
\tilde{\sigma}=\frac{\sigma}{\sqrt{3H_0^2\Omega_{\Lambda}}},
\eeq
and hence we can write the Sachs equations in the form
\beqa
d''_A+d'_A\frac{E'}{E}&=&\left(\frac{4}{9}\frac{\phi_{00}}{E^2H_0^2\Omega_{\Lambda}}-\frac{4}{3}\frac{|\tilde{\sigma}|^2}{E^2}\right)d_A,\\
\tilde{\sigma}'&+&2\frac{d'_A}{d_A}\tilde{\sigma}=\frac{2}{3\sqrt{3}E H_0^2\Omega_{\Lambda}}\Psi_0,
\eeqa
where, again, a dash denotes differentiation with respect to the time variable $\tau$. To calculate the Ricci focusing term $\phi_{00}$ and the Weyl focusing term $\Psi_0$, we need the form of the complex null tetrad. We shall firstly consider the case where the photon travels along the $z$-axis and then generalise the result to light rays travelling in any direction. For a light ray travelling along the $z$-axis we find
\beq
l^a=E(1,0,0,\frac{1}{SZ}),
\eeq
and
\beq
m^a=\frac{1}{\sqrt{2}}(0,\frac{1}{S},\frac{-i}{S},0),
\eeq
where $E$, again, is the energy of the photon. This tetrad is very similar to the one derived in \citet{MeuBru11-1}, except for the $E$ factor in $l^a$ and so, using this tetrad, we find the only non-zero Weyl scalar to be\footnote{In doing this, we have chosen a null tetrad which is especially adapted to the metric: having $\Psi_2$ as the only non-zero Weyl scalar is characteristic of the Petrov type D of our space-time, see \citet{MeuBru11-1} for more details. The fact that a metric of the form we are using is of Pertov type D was first shown by \citet{BarRow89}.} $\Psi_2$. Hence, for light bundles along the $z$-axis, we find the two focusing terms
\beq
\phi_{00}=-\frac{1}{2}E^2\rho,
\eeq
and
\beq
\Psi_0=0,
\eeq
while
\beq\label{eq:psi2}
\Psi_2=\frac{1}{6}\bar{\rho}\delta
\eeq
This brings the Sachs optical equations along the $z$-axis into the form
\beqa
d''_A+d'_A\frac{E'}{E}&=&\left(-\frac{2}{9}\frac{\rho}{H_0^2\Omega_{\Lambda}}-\frac{4}{3}\frac{|\tilde{\sigma}|^2}{E^2}\right)d_A, \label{eq:sachsd}\\
\tilde{\sigma}'&+&2\frac{d'_A}{d_A}\tilde{\sigma}=0.\label{eq:sachss}
\eeqa
This system of equations should be integrated from today back into the past and so we need to set initial conditions today, say $\tau\so$ and we set $\tilde{\sigma}|\so=0$, $d_A|\so=0$ and $d'_A|\so=-2/(3E|\so H_0\sqrt{\Omega_{\Lambda}})$. Given these initial conditions, it is apparent from Eq.\ (\ref{eq:sachss}) that $\tilde{\sigma}|\so=0$ implies the trivial solution $\tilde{\sigma}=0$ and hence we only have to consider Eq.\ (\ref{eq:sachsd}) and the initial conditions associated with it. Essentially this means that along the $z$-axis the Weyl focusing is zero and hence the light bundles do not experience any shear, however, the Weyl focusing is an effect we are interested in, as it might have non-negligible effects on the angular diameter distance. Hence, we will now generalise the above treatment to be able to consider light bundles that do not travel along the $z$-axis.

We are interested in the case where $\gamma=\omega=0$, since these two functions only displace the center of deviations, and therefore in this special case, our model displays an axial symmetry about the $z$-axis, see \citet{MeuBru11-1}. This implies that considering light rays in the $y$-$z$-plane is completely general, as one could always do a rotation about the $z$-axis without changing the metric but making the tangent vector point out of the $y$-$z$-plane. We name the angle that the tangent vector $l^a$ subtends with the $z$-axis $\alpha$ and all quantities in the rotated system are denoted with a tilde. The null tetrad in the rotated frame then takes the form
\beq\label{eq:tet1}
\tilde{l}^a=E(1,0,\frac{\sin(\alpha)}{S},\frac{\cos(\alpha)}{SZ}),
\eeq
and
\beq\label{eq:tet2}
\tilde{m}^a=\frac{1}{\sqrt{2}}(0,\frac{1}{S},\frac{-i\cos(\alpha)}{S},\frac{-i\sin(\alpha)}{SZ}).
\eeq
Since we are dealing with pure $\Lambda$CDM, dust and a cosmological constant, it follows from EFEs that the Ricci focusing term does not depend on rotations in the basis vectors and so we find that
\beq
\tilde{\phi}_{00}=\phi_{00}=-\frac{1}{2}E^2\rho.
\eeq
However, deriving the Weyl focusing term in the rotated frame is not as straight forward and we need to consider the effect of rotations in the complex null tetrad on the Weyl scalars. In the complex null tetrad constructed for light rays travelling along the $z$-axis, we found that the only non-zero Weyl scalar was $\Psi_2$, Eq.\ (\ref{eq:psi2}). Rotating the original complex null tetrad to coincide with the the physical situation of light propagation at an angle $\alpha$ with the $z$-axis requires four separate canonical rotations in the complex null tetrad, see Appendix \ref{sec:trans} for the details. These rotations have the effect of making all five Weyl scalars non-zero in general, but expressible in terms of the original $\Psi_2$, in particular, we find
\beq\label{eq:psitilde}
\tilde{\Psi}_0=-3\sin^2(\alpha)E^2\Psi_2=-\frac{1}{2}\sin^2(\alpha)E^2\bar{\rho}\delta,
\eeq
where $\bar{\rho}$ is the background density and $\delta$ is the density deviation. Therefore, we find the general Sachs optical equations for our space-time
\beqa
d''_A+d'_A\frac{E'}{E}&=&\left(-\frac{2}{9}\frac{\rho}{H_0^2\Omega_{\Lambda}}-\frac{4}{3}\frac{|\tilde{\sigma}|^2}{E^2}\right)d_A, \label{eq:sachsdgen}\\
\tilde{\sigma}'&+&2\frac{d'_A}{d_A}\tilde{\sigma}=-\frac{1}{3\sqrt{3}}\frac{\sin^2(\alpha)E}{H_0^2\Omega_{\Lambda}}\bar{\rho}\delta, \label{eq:sachssgen}
\eeqa
where we choose initial conditions again as $\tilde{\sigma}|\so=0$, $d_A|\so=0$ and $d'_A|\so=-2/(3E|\so H_0\sqrt{\Omega_{\Lambda}})$ and integrate into the past. Clearly, choosing $\tilde{\sigma}|\so=0$ here does not imply the trivial solution for $\tilde{\sigma}$, since the Weyl focusing term is non-zero in general.

\section{Results of the light tracing}\label{sec:res}
In this section, we would like to present how the results of the light tracing we performed in our model differ from the standard FLRW results commonly used. To find the position and redshift of the bundle of light rays, we need to integrate Eqs.\ (\ref{eq:odeE})-(\ref{eq:oder}) in general, whereas, to find the angular diameter distance and shear of the bundle, we need to integrate Eqs.\ (\ref{eq:sachsdgen}) and (\ref{eq:sachssgen}). Inspecting the last two equations, we find that they are coupled to the geodesic equations and hence, we need to solve all six differential equations simultaneously. From the angular diameter distance, we can find the luminosity distance, using Eq.\ (\ref{eq:Eth}), but we would also like to compare the distance modulus we find to the standard FLRW result and therefore we use
\beq
\Delta d_M=d_M-d_M^{FLRW}=5\log_{10}\left( \frac{d_L}{d_L^{FLRW}}\right),
\eeq
where $d_M$ is the distance modulus and $d_L$ is the luminosity distance, where a superscript FLRW denotes the same quantity in the FLRW background. To compare the angular diameter distance and redshift in our model to the standard FLRW result, we choose the definitions
\beq
\Delta z=\frac{z^{IN}-z^{FLRW}}{z^{FLRW}},
\eeq
and
\beq
\Delta d_A=\frac{d_A-d_A^{FLRW}}{d_A^{FLRW}},
\eeq
which should give the reader an intuitive idea of what the fractional difference is between the results we derive here and the commonly used FLRW values.

Having derived the geodesic equations and Sachs optical equations and having defined quantities to analyse our results, we have to consider which physical situations we would like to model. For a detailed discussion of which matter distributions are possible and a discussion of singularities, see the relevant sections in \citet{MeuBru11-1}. Here, however, we would like to mention the main points that characterise the matter distributions we can model. In this paper, we concentrate on the $\gamma=\omega=0$ case, which reduces the freedom of the model to one function, the space distribution of the growing mode $\beta_+$, which gives us the freedom to set the initial matter distribution along the $z$-axis, whereas we do not have any freedom to set the distributions along the $x$- and $y$-axis. If we only consider under-densities, ($\beta_+>0$) we do not find singularities in any space-time point. However, as long as $\beta_+$ is negative in some region, which corresponds to an over-density in the same region, pancakes can eventually form in the model - as also expected from Newtonian gravitational collapse. What is important for the analysis presented here is that for over-densities which have not yet collapsed by today (which is what we are interested in), we can, at any time in the past, find a region around the $z$-axis which is free from singularities and therefore we can perform light tracing in those regions.

\subsection{Single mode density distributions}\label{sec:res1}
As we explained in Sec.\ \ref{sec:sol}, the linearity of Eq.\ (\ref{eq:diffFhom}) allows for the validity of a superposition principle for the metric function $F$. In addition, we only consider the growing mode whose spatial distribution is encoded in the function $\beta_+(z)$. Therefore, in this section, we first look at harmonic distributions of matter along the line of sight, i.e.\ single mode sinusoidal distributions.

\begin{figure}
\begin{center}
$\begin{array}{rr}
\includegraphics[width=0.4\textwidth]{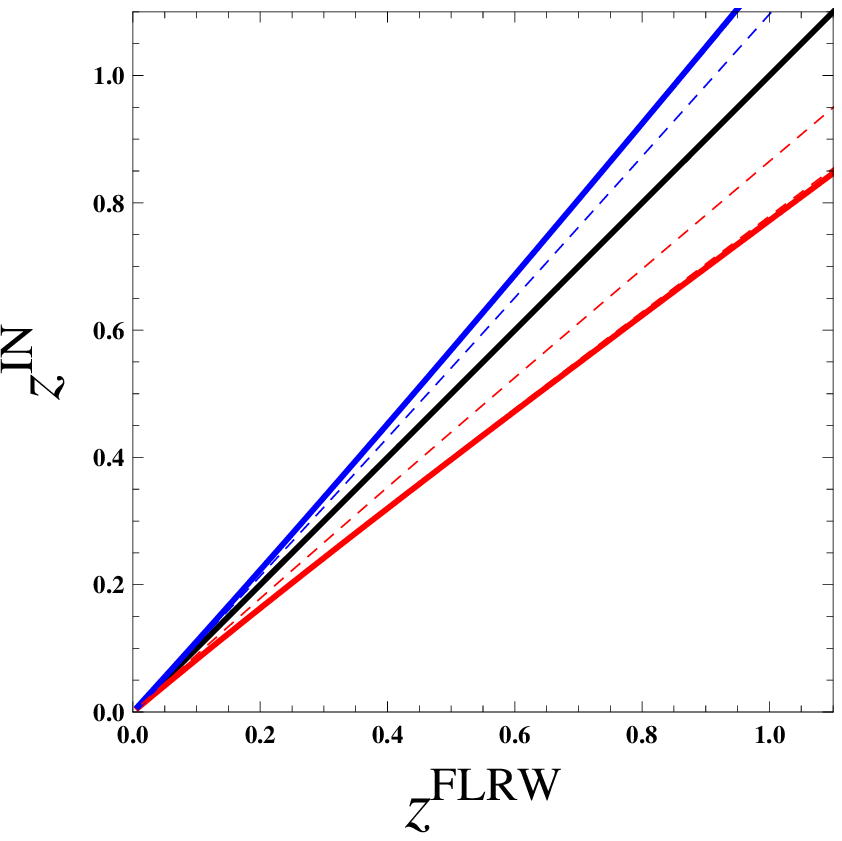} \\
\includegraphics[width=0.4\textwidth]{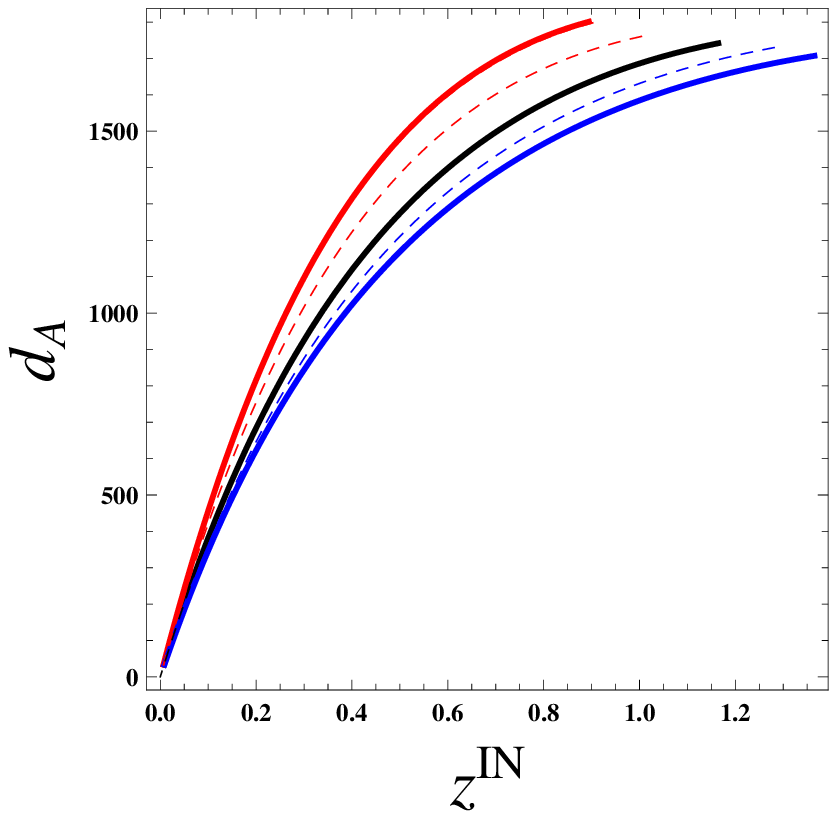}  \\
\end{array}$
\end{center}
\caption{Redshift $z^{IN}$ found in our model, as compared to the redshift $z^{FLRW}$ the same object would have in an FLRW model, top panel, and angular diameter distance $d_A$ as a function of observed redshift $z^{IN}$, bottom panel. We are considering only over-densities (top, red lines), only under-densities (bottom, blue lines) and compensated density distributions (middle, black lines). The deviations are periodic on scales of $8$ Mpc along the $z$-axis, the solid lines are for on-axis light rays and the dashed lines for off-axis rays (corresponding to $10$ and $40$ degree deviations). The dashed lines for the compensated case are not visible here, as they are not distinguishable form the on-axis case at this resolution.}
\label{fig:dzda}
\end{figure}

\begin{figure}
\includegraphics[width=0.45\textwidth]{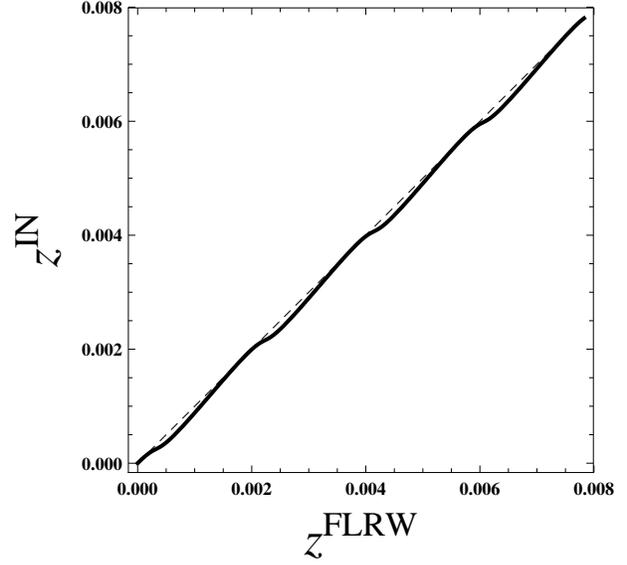}
\caption{Redshift found in our model $z^{IN}$ versus the redshift an object at the same distance from us would have in an FLRW model $z^{FLRW}$. The solid line is the redshift found from the light tracing for a compensated density profile, which corresponds to $\delta \approx 1$ today, periodic on a scale of $8$ Mpc today. We are only plotting very small redshifts here, so the deviations found are visible. The solid line corresponds to the redshifts we find from the light tracing and the dashed line corresponds to the FLRW values, plotted for reference. This is a zoom in to the black line on the top panel of Fig.\ \ref{fig:dzda}.}
\label{fig:zzoom}
\end{figure}

The first question that comes to mind is whether to consider a distribution of over-densities, under-densities or a combination thereof. As a first analysis, we would like to see what the effect of either of those three choices is and hence in Fig.\ \ref{fig:dzda} we present the redshift and the angular diameter distance $d_A$ obtained from a model with only over-densities (red lines), only under-densities (blue lines) and compensated density profiles (black lines). The solid lines are for light rays along the $z$-axis and the dashed lines correspond to light rays which travel at an angle of $10$ and $40$ degrees off the $z$-axis, the $40$ degrees lines are always the ones further away from the respective solid line. No deviation between the different angles is visible for the black line as no difference from the FLRW curves are visible at this resolution for any angle in the compensated case. In Fig.\ \ref{fig:dzda} we have chosen all inhomogeneities to be periodic on a scale of $8$ Mpc, the initial conditions have been chosen such that at early times, $\delta_{over}\propto 1-\cos(\frac{2\pi z}{8Mpc})$, $\delta_{under}\propto \cos(\frac{2\pi z}{8Mpc})-1$ and $\delta_{comp}\propto \cos(\frac{2\pi z}{8Mpc})$, where $\delta_{comp}$ stands for a density perturbation which is compensated on the above mentioned $8$ Mpc scale along the $z$-axis. The amplitude of the over-densities in the compensated and only over-density cases correspond to $\delta\approx1$ today, whereas the under-densities grow to voids of $\delta\approx-0.3$ today. The figures show significant deviations in the redshift and distance measure, if we only consider over- or under-densities along the line of sight. One might get the impression here that for the compensated case there are no deviations from the FLRW values, therefore in Fig.\ \ref{fig:zzoom} we have plotted a zoom in on the very small redshift range of Fig.\ \ref{fig:dzda} for the compensated case only and a periodic deviations is clearly visible.

\begin{figure}
\begin{center}
$\begin{array}{llll}
\includegraphics[width=0.4\textwidth]{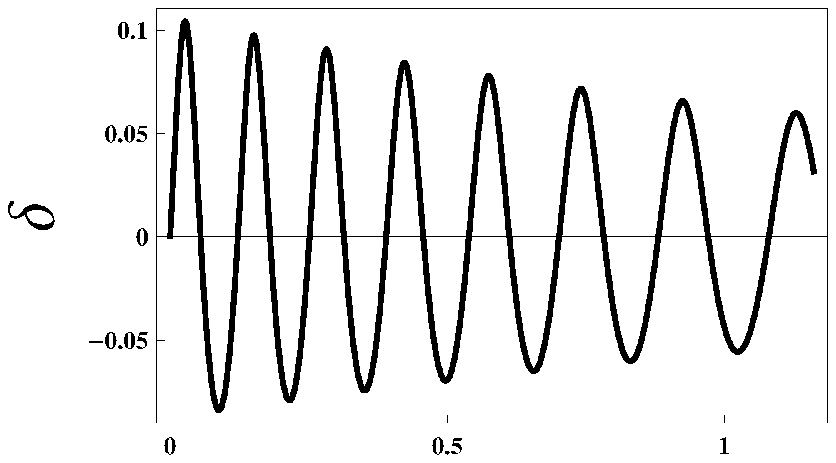} \\
\includegraphics[width=0.4\textwidth]{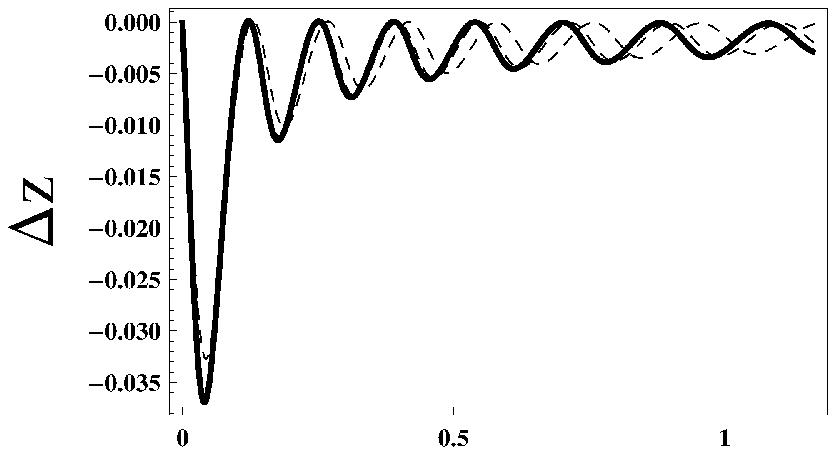}  \\
\includegraphics[width=0.4\textwidth]{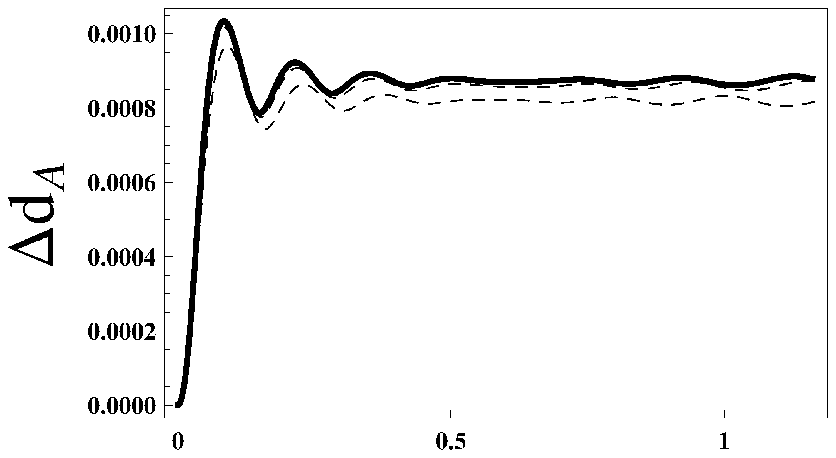} \\
\includegraphics[width=0.4\textwidth]{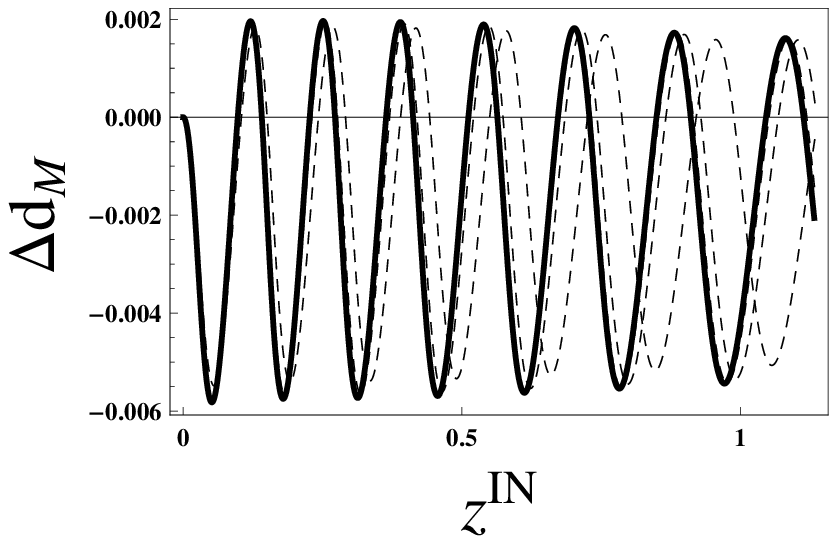}  \\
\end{array}$
\end{center}
\caption{The effect of inhomogeneities with $\delta\approx0.1$ today, periodic on a scale of $500$ Mpc. The solid line corresponds to light rays travelling along the $z$-axis and the dashed lines to light rays travelling at $5$, $10$ and $20$ degrees off-axis respectively.}
\label{fig:dev500}
\end{figure}

\begin{figure}
\begin{center}
$\begin{array}{llll}
\includegraphics[width=0.4\textwidth]{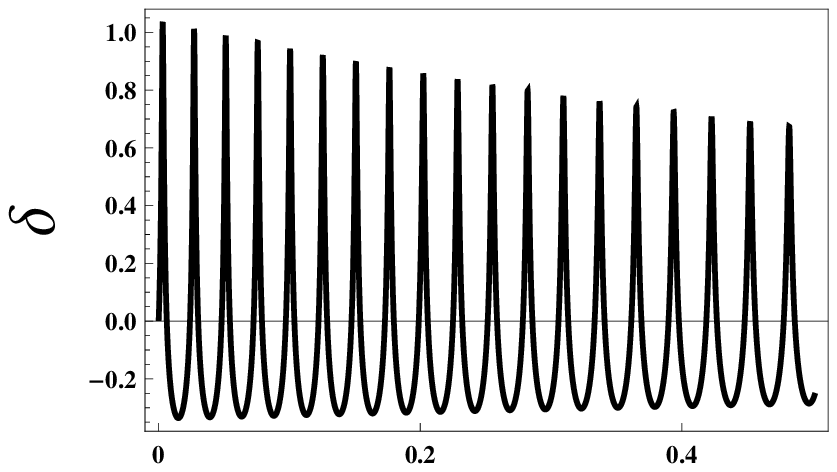} \\
\includegraphics[width=0.4\textwidth]{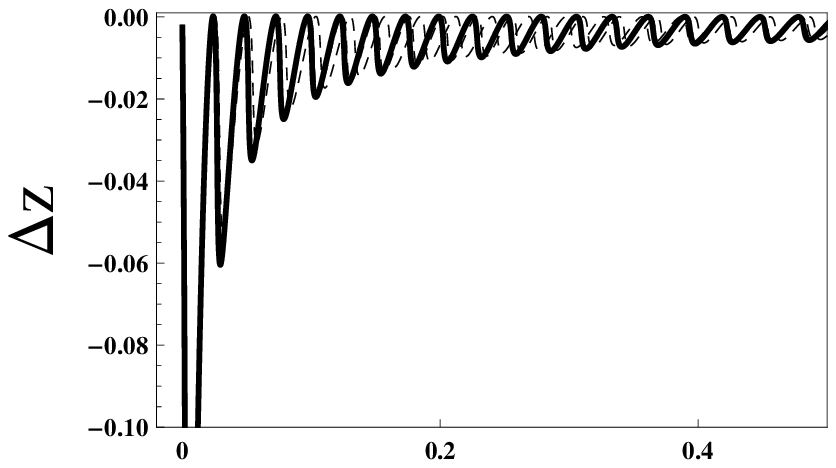}  \\
\includegraphics[width=0.4\textwidth]{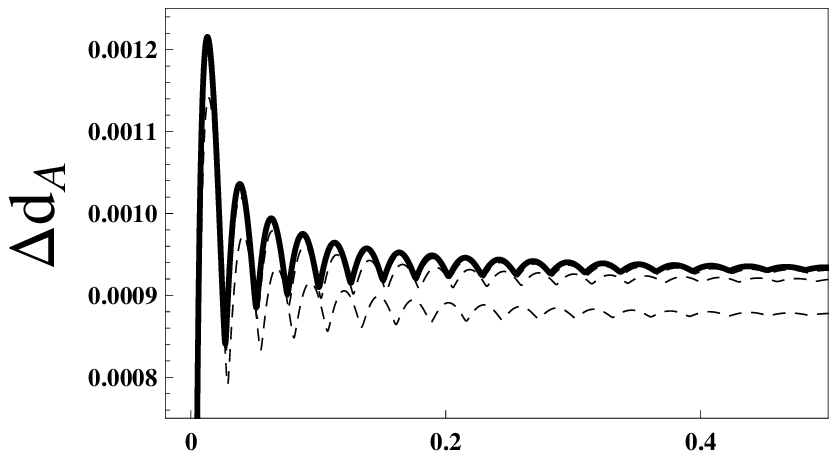} \\
\includegraphics[width=0.4\textwidth]{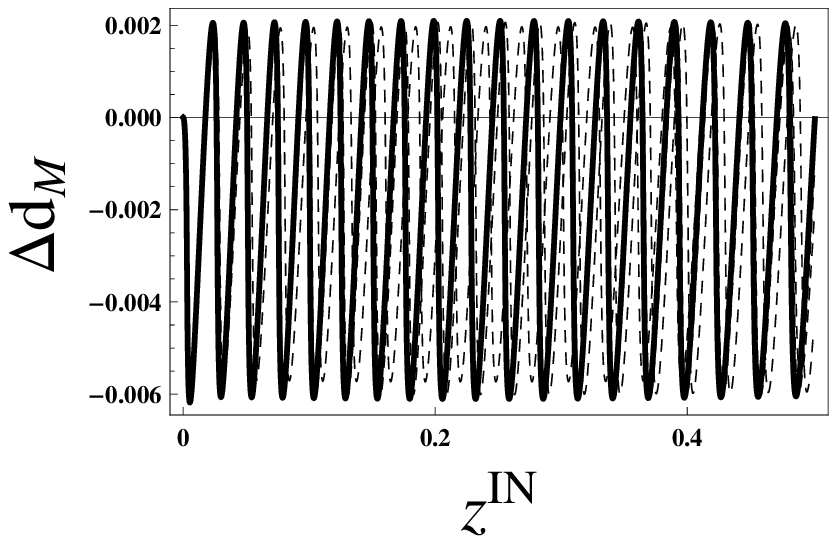}  \\
\end{array}$
\end{center}
\caption{The effect of inhomogeneities with $\delta\approx1$ today, periodic on a scale of $100$ Mpc. The solid line corresponds to light rays travelling along the $z$-axis and the dashed lines to light rays travelling at $5$, $10$ and $20$ degrees off-axis respectively.}
\label{fig:dev100}
\end{figure}

\begin{figure}
\begin{center}
$\begin{array}{llll}
\includegraphics[width=0.4\textwidth]{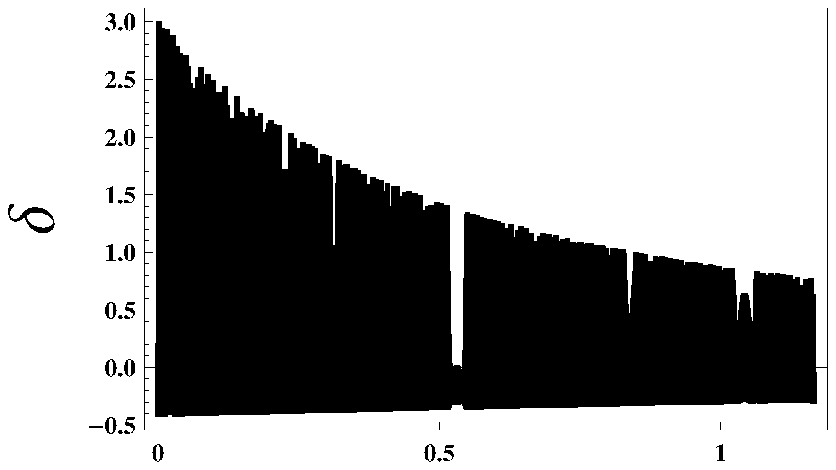} \\
\includegraphics[width=0.4\textwidth]{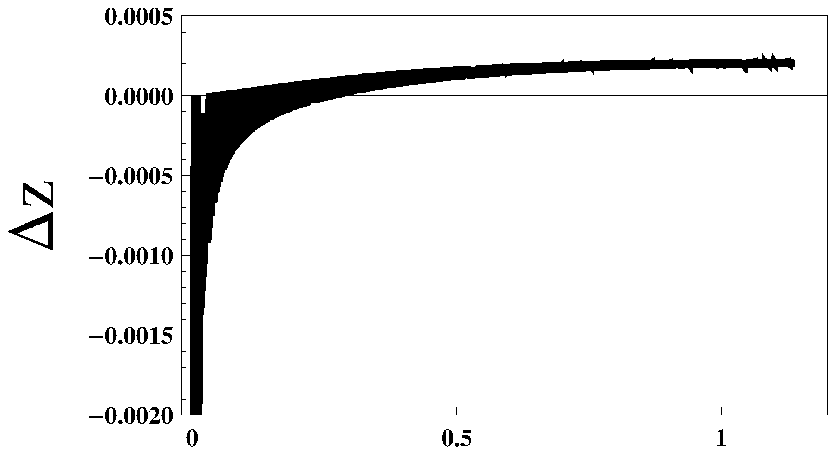}  \\
\includegraphics[width=0.4\textwidth]{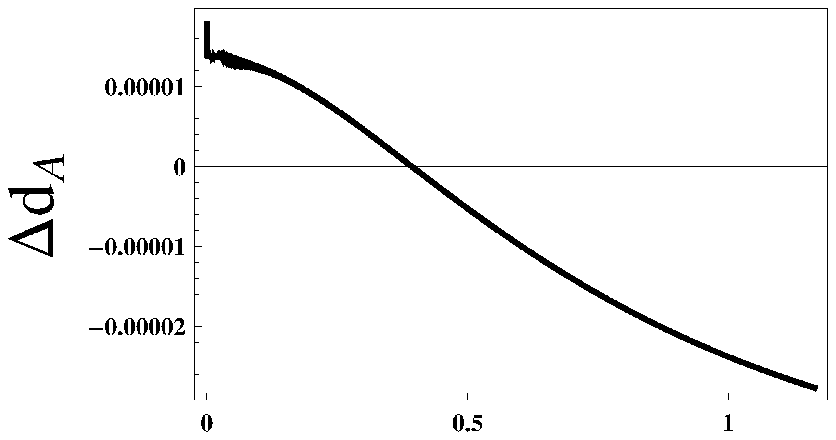} \\
\includegraphics[width=0.4\textwidth]{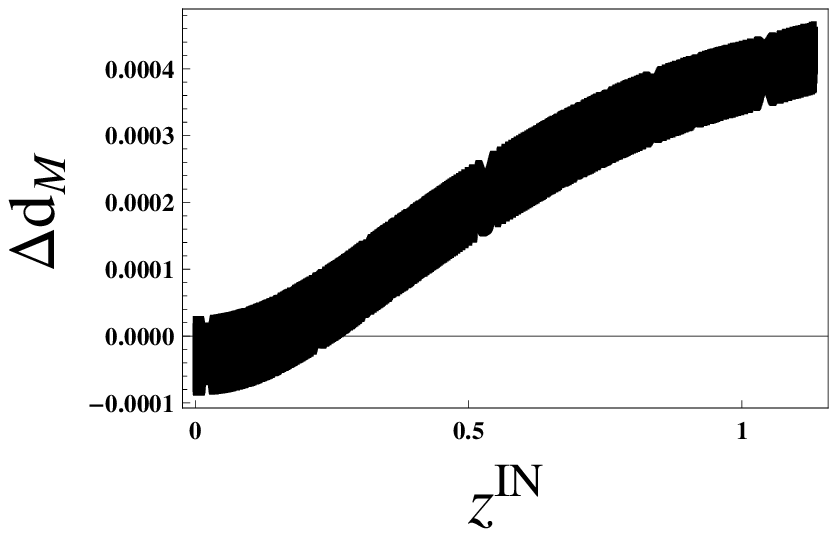}  \\
\end{array}$
\end{center}
\caption{The effect of inhomogeneities with $\delta\approx3$ today, periodic on a scale of $1$ Mpc. Here we are only showing the results from light tracing along the $z$-axis for clarity. Two features are visible in the lower three panels, the thickness of the lines shows the effect of each single oscillation in the density profile, whereas the trends of these thick lines shows the overall integrated effect of having large amplitude but small scale density inhomogeneities. }
\label{fig:dev1}
\end{figure}

However, when we make actual observations in the Universe, we generally assume that we observe along typical lines of sight and that the density deviation along this line of sight should average to zero, or at least we expect that the ensemble average of the density deviation along many lines of sight in different directions should average to zero. Hence, we would like to consider in the following matter distributions which we can show to average to zero over some characteristic distance. This is automatically achieved with the harmonic sinusoidal distribution we are dealing with which clearly implies a zero average over one period of the function. Figs.\ \ref{fig:dev500}, \ref{fig:dev100} and \ref{fig:dev1} show the density deviation $\delta$, redshift deviation $\Delta z$, angular diameter distance deviation $\Delta d_A$ and the distance modulus deviation $\Delta d_M$ for density deviations on different scales and of different amplitudes. On all plots, the solid lines correspond to light rays which travelled along the $z$-axis and dashed lines correspond to off-axis light rays. The off-axis light rays were directed at angles of $5$, $10$ and $20$ degrees from the $z$-axis. To distinguish the lines, one can assume that the ones that deviate from the solid line the most are the ones sent at an $20$ degree angle and the ones sent at $5$ degrees are hardly distinguishable from the solid lines. We are considering here density deviations on different scales, ranging from $8$ Mpc to $500$ Mpc. However, we do not expect density deviations to be of the same amplitude across all these scales: in general, while the Universe is very inhomogeneous on small scales, observations support the idea that deviations from homogeneity get smaller and smaller on larger scales, where homogeneity is reached at some point \citep{Sar09}. Generally, we will be considering larger amplitude density deviations on smaller scales and smaller amplitudes on larger scales. Despite this, interestingly we find deviations in the distance measures and redshift to be mostly affected by the larger scale density deviations. Considering larger amplitude density deviations on larger scale would simply amplify this effect but make the density distributions less realistic. Note that in Fig.\ \ref{fig:dev100} we are using a smaller range in redshift than in the other two figures to make it easier to tell apart features of the off- and on-axis lines.

Looking at Figs.\ \ref{fig:dev500}, \ref{fig:dev100} and \ref{fig:dev1}, we can see deviations from the FLRW background values in all the quantities we display, however all these deviations seem to be below the $1\%$ level, given the conservative assumptions we made. 

We would like to understand what term in the Sachs equations, Eqs.\ (\ref{eq:sachs1}) and (\ref{eq:sachs2}),  is the main cause of these deviations and hence we consider the non-FLRW parts of the Ricci focusing, the Weyl focusing and the shear resulting from Eqs.\ (\ref{eq:sachsdgen}) and (\ref{eq:sachssgen}). For the Ricci focusing, we will introduce the variable
\beq\label{eq:dphi}
\Delta \phi=\phi_{00}-\phi_{00}^{FLRW}=-\frac{1}{2}\left(E^2\rho-\bar{E}^2\bar{\rho} \right),
\eeq
where $\bar{E}$ is the photon energy in the background FLRW model. For the Weyl focusing, there is no background contribution, hence, we just have to consider the Weyl scalar $\Psi_0$; for the contribution of the shear, we will consider $|\sigma|^2$ as this is the term present in the Sachs equations, and again, $\sigma$ vanishes in the background given our initial conditions. 

\begin{figure}
\begin{center}
$\begin{array}{rrr}
\includegraphics[width=0.45\textwidth]{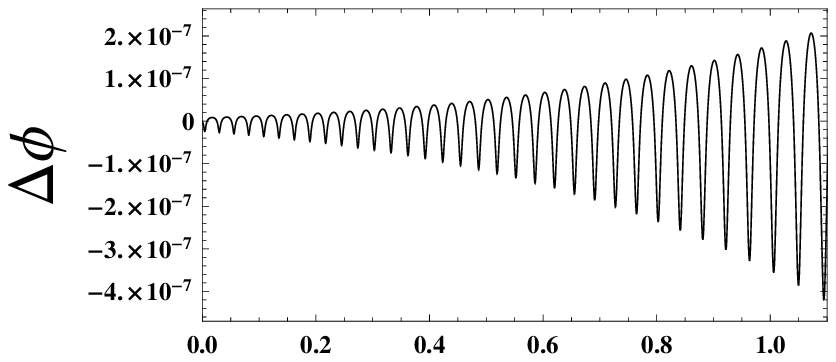} \\
\includegraphics[width=0.45\textwidth]{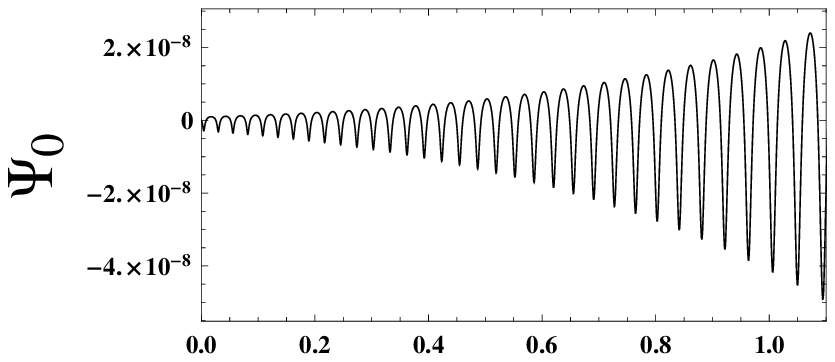}  \\
\includegraphics[width=0.42\textwidth]{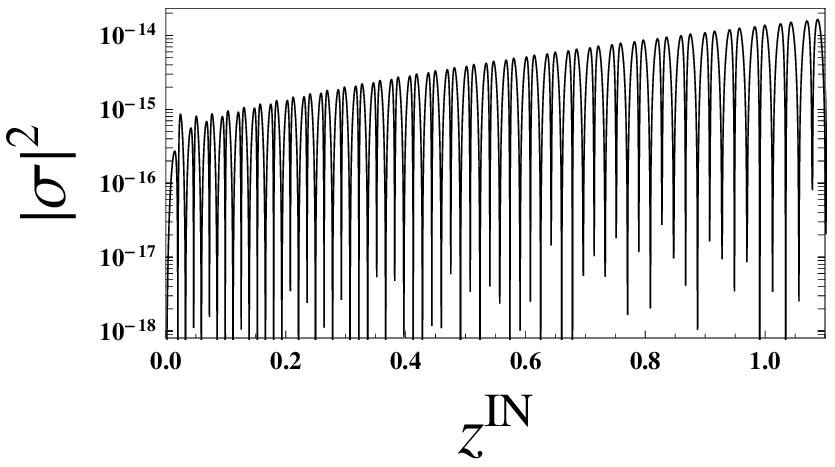} \\
\end{array}$
\end{center}
\caption{Plot of the non-FLRW contributions in the Sachs equations as a function of redshift. Here we have used compensated density deviations on scales of $100$ Mpc which would have an amplitude of $\delta\approx 1$ today and the light tracing was performed at an $20$ degree angle with the $z$-axis. Where $\Delta\phi=\phi_{00}-\phi_{00}^{FLRW}$, $\Psi_0$ is the zeroth Weyl scalar and $\sigma$ is the shear of the light bundle.}
\label{fig:focus}
\end{figure}

In Fig.\ \ref{fig:focus} we compare the above mentioned variables for a representative matter distribution: we choose $\delta$ periodic on scales of $100$ Mpc along the $z$-axis with an amplitude of $\delta\approx1$ today. We have performed the integration for light rays which travel at an angle of $20$ degrees with the $z$-axis. From the plots we can see that the Ricci focusing deviation dominates over the Weyl focusing. Although from Eq.\ (\ref{eq:dphi}) we see that $\Delta\phi\approx \bar{E}\bar{\rho}\delta$ so that the source of both the Ricci and Weyl focusing is $\bar{\rho}\delta$, the Weyl focusing is strongly suppressed by the $\sin^2(\alpha)$ factor in Eq.\ (\ref{eq:psitilde}). Finally, we find that $|\sigma|^2$ is much smaller than the two focusing terms, so that it gives a negligible contribution in Eq.\ (\ref{eq:sachs1}) and, consequently, to the deviations from FLRW in the distance measures. 

\subsection{Mode coupling and its effects}\label{sec:res2}
The above results show that single mode density deviations do not have a large effect on the redshift, angular diameter distance and distance modulus for the compensated profiles we considered. From this we cannot conclude though that a density profile, where the metric function $F$ is the sum of many modes, has a small effect as well, since the structures in our model grow non-linearly. In other words, an initial deviation consisting of the superposition of two modes in $F$ may excite many different modes in $\delta$ during its non-linear growth and the resultant density profile might have completely different effects on the redshift and the distance modulus. The real Universe clearly does not consist of only one wavelength deviation and therefore the step of including several modes should make our analysis more realistic.

\begin{figure}
\begin{center}
$\begin{array}{rrr}
\includegraphics[width=0.4\textwidth]{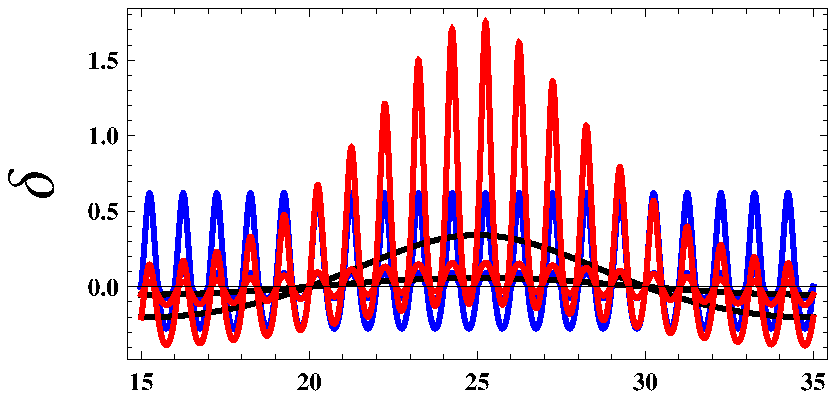} \\
\includegraphics[width=0.4\textwidth]{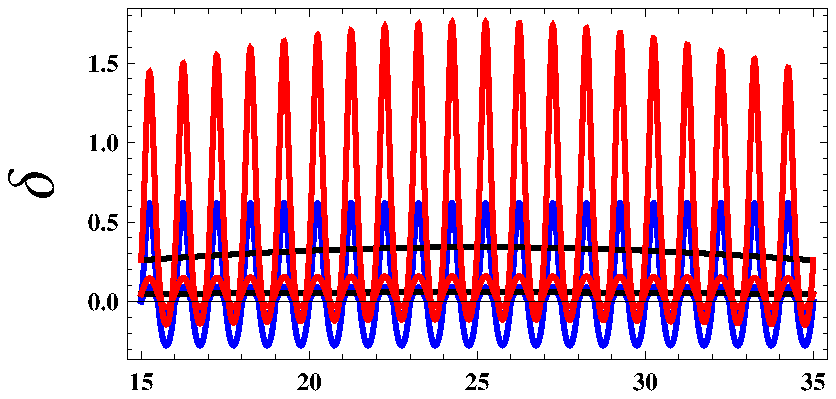}  \\
\includegraphics[width=0.39\textwidth]{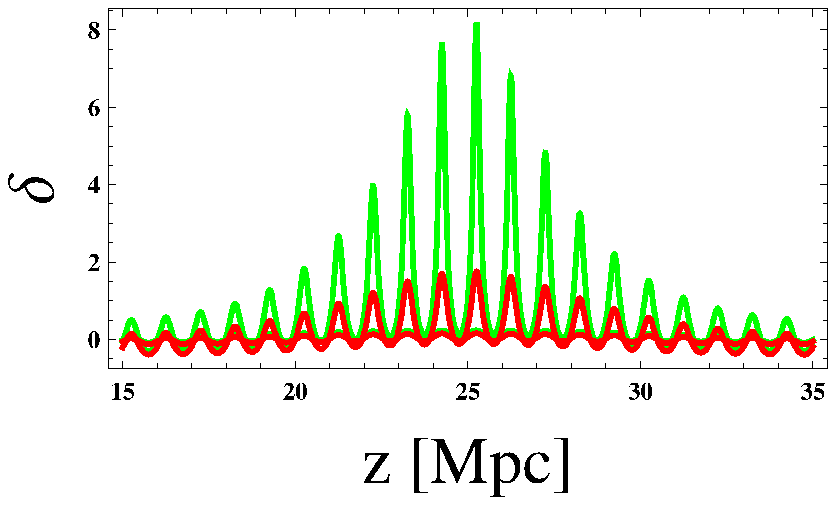} \\
\end{array}$
\end{center}
\caption{Illustration of how different modes interact in the density profile. Each profile is plotted at a redshift of $z^{FLRW}=5$ (smaller amplitude curves) and today. The top panel demonstrates the interaction of a mode that is periodic on $1$ Mpc, blue curves, and a mode periodic on $20$ Mpc, black curves, and how the two modes interact, red curves, if superimposed as initial conditions. The same is shown in the middle panel for modes periodic on $1$ Mpc, blue curves, and $100$ Mpc, black curves, and their interaction, red curves. In the bottom panel, we show how an initial superposition of the $1$ Mpc and $20$ Mpc modes, red curves, behave compared to an initial superposition of all three, $1$ Mpc, $20$ Mpc and $100$ Mpc modes, green curves.}
\label{fig:deltas}
\end{figure}

In general in our model we find that the growth of single mode density deviations does not depend on their wavelength, that is, if we only had deviations periodic on a $1$ Mpc scale, they would grow to the same amplitude as if we only had deviations on a scale of $100$ Mpc. If we had an initial small density deviation which is a superposition of those two modes in $F$, the growth of $\delta$ would be quite different to the individual modes. In Fig.\ \ref{fig:deltas} we show how modes on scales of $1$ Mpc, $20$ Mpc and $100$ Mpc interact. From these plots one can see how peaks in the long wavelength perturbation cause the short wavelength perturbations to grow non-linearly and it becomes clear that the profiles are not superposition of the individual modes any more. More precisely, in the top panel of Fig.\ \ref{fig:deltas} we show the growth of a $1$ Mpc mode together with a $20$ Mpc mode and their interaction. It is clear that the growth of the short wavelength mode gets an extra non-linear kick from growing on top of the larger scale mode. On the other hand, peaks of the short scale mode that grow in the voids of the larger scale mode are depressed.  The same qualitative behaviour can be observed in the middle panel of Fig.\ \ref{fig:deltas}, where we now consider the $1$ Mpc mode together with a $100$ Mpc mode, plotting on the same length scale as the top panel. In this case the peak of the larger scale mode (in the middle of the figure) is very broad and the short wavelength mode is growing almost as if on top of a different background. However, non-linearity is again important and the peaks are much higher than they would be if simply raised by this ``new background". Finally, in the bottom panel we show the effect of adding the larger $100$ Mpc mode to the deviations in the top panel. The red profile for the coupled $1$ Mpc and $20$ Mpc modes of the top panel is shown again in red in the bottom panel. The green profile shows the effect of coupling the three modes together. This bottom plot in Fig.\ \ref{fig:deltas} therefore shows that even adding a small amplitude ($\delta\approx0.3$ today) $100$ Mpc mode to the $1$ Mpc and $20$ Mpc allows the peaks on the shorter scale to grow much stronger, more than a factor of $4$ than without. 

\begin{figure}
\begin{center}
$\begin{array}{llll}
\includegraphics[width=0.4\textwidth]{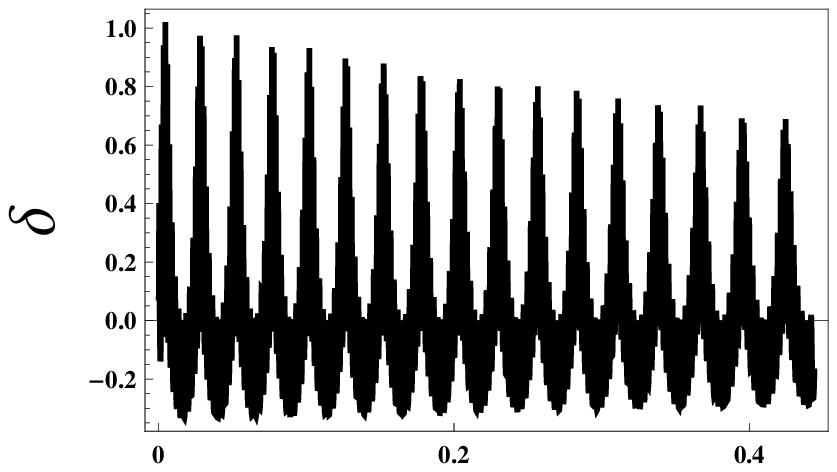} \\
\includegraphics[width=0.4\textwidth]{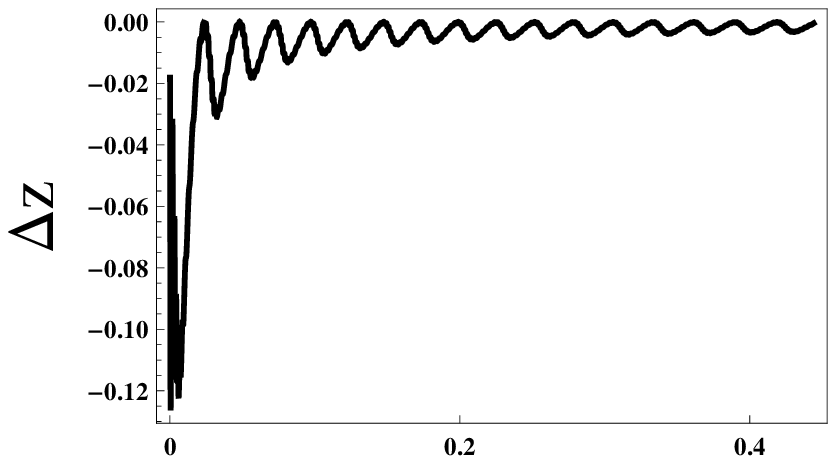}  \\
\includegraphics[width=0.4\textwidth]{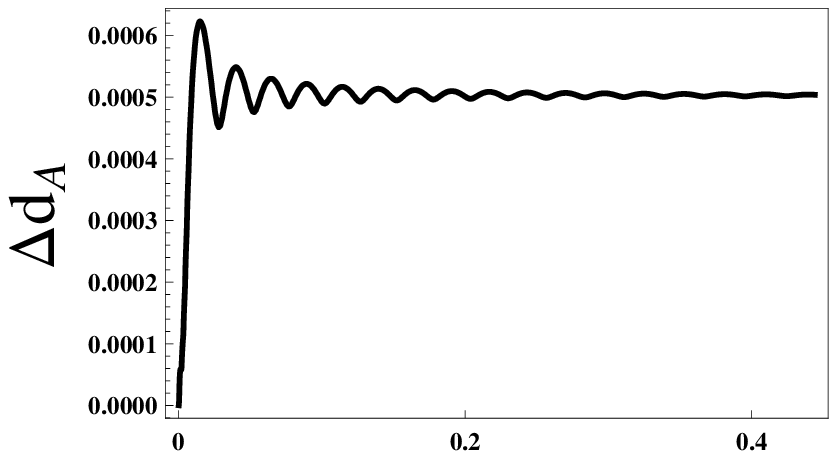} \\
\includegraphics[width=0.4\textwidth]{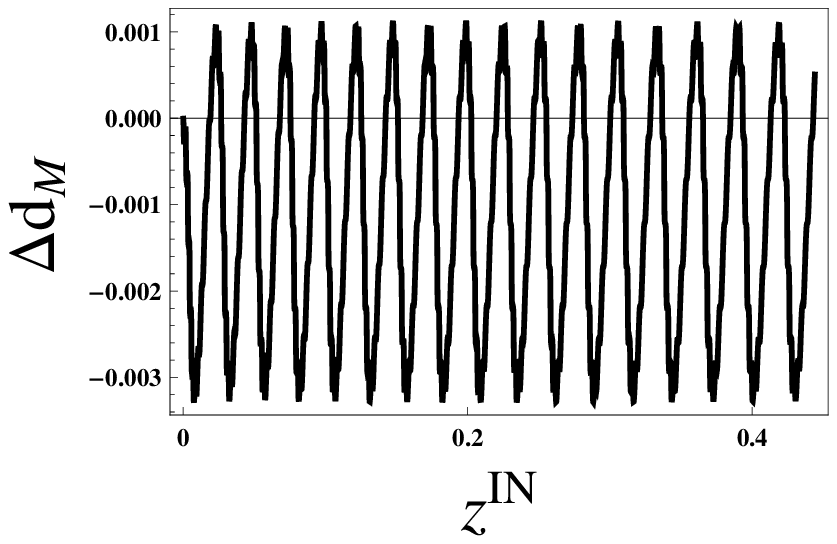}  \\
\end{array}$
\end{center}
\caption{The effect of inhomogeneities with $\delta\approx1$ today, with the initial condition being a superposition of two modes, one period on scales of $1$ Mpc and the other on $100$ Mpc. The initial amplitude of the two different perturbations was chosen to be the same. Here we are only showing the results from light tracing along the $z$-axis for clarity.}
\label{fig:dev1100}
\end{figure}

To investigate the effect of the mode coupling on the redshift and distances, in Fig.\ \ref{fig:dev1100} we show how two coupled modes, one $1$ Mpc mode and one $100$ Mpc mode, affect the redshift and distances. The two modes have been chosen to be of equal initial amplitude and combine to result in structure of $\delta \approx 1$ today. The amplitude of deviations in redshift, angular diameter distance and distance modulus are completely dominated by the effects of the long wavelength, $100$ Mpc deviation. This analysis has been done for many more pairs of modes, from $1$ Mpc to $500$ Mpc, and the results always seem to be dominated by the long wavelength modes. The fact that the main effects in redshift and therefore in distances is dominated by the larger scales inhomogeneities should not come as a surprise. The basic mechanism at work here is the same as in the Rees-Sciama and the integrated Sachs-Wolfe effect. In an expanding universe photons travel through dynamical inhomogeneities. When the characteristic scale of the inhomogeneity is negligible compared to the Hubble radius, the effect of the expansion is negligible and therefore a photon will come out of the inhomogeneity with the same energy that it had when it entered it. Instead, in going through a large scale inhomogeneity, photons have to go through a different potential well when they come out of the inhomogeneity than when they were entering it, changing their energy in the process.

\subsection{Peaks and voids of arbitrary profile}\label{sec:res3}
\begin{figure}
\begin{center}
$\begin{array}{llll}
\includegraphics[width=0.4\textwidth]{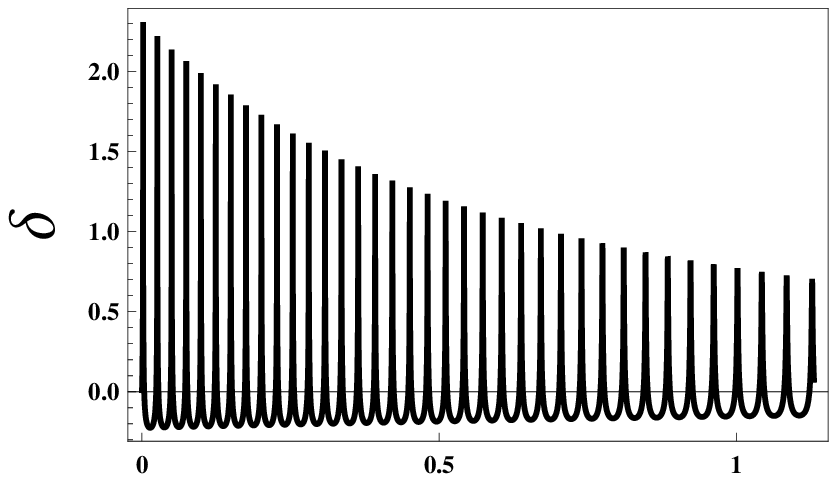} \\
\includegraphics[width=0.4\textwidth]{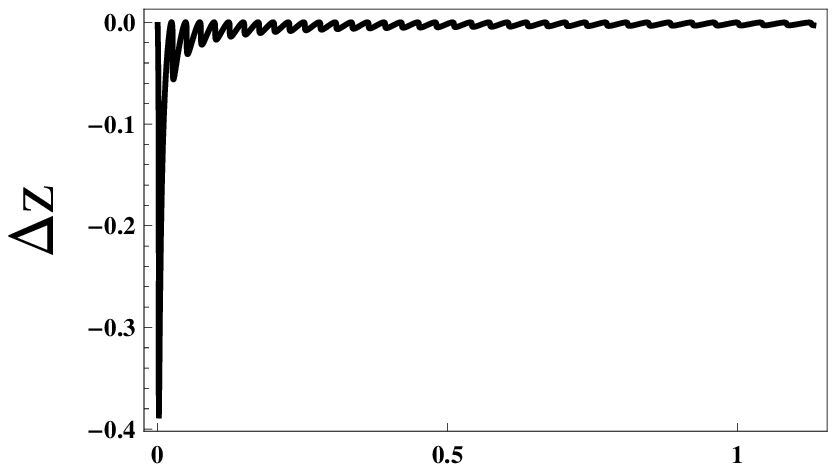}  \\
\includegraphics[width=0.4\textwidth]{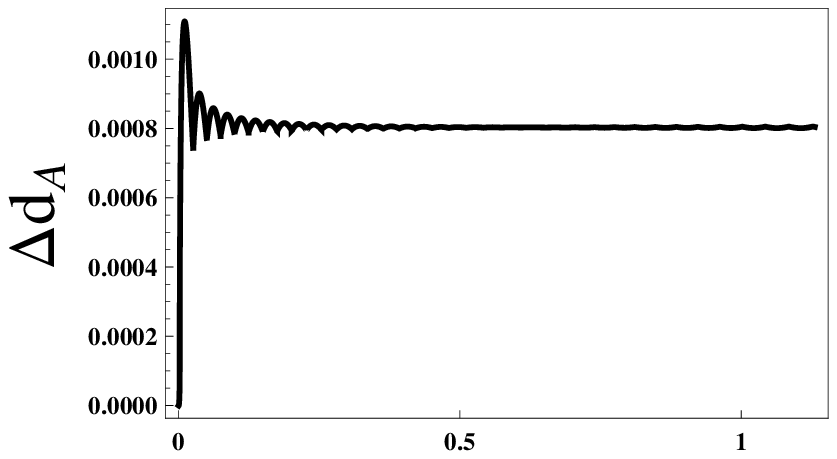} \\
\includegraphics[width=0.4\textwidth]{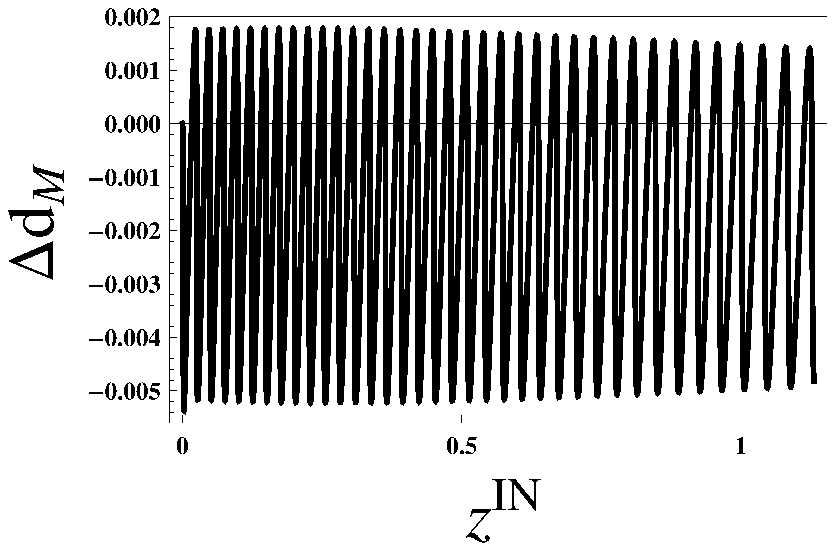}  \\
\end{array}$
\end{center}
\caption{The effect of inhomogeneities with $\delta\approx2.3$ today, with the choice of the initial profile shown in Eq.\ (\ref{eq:cosh}). Here we are only showing the results from light tracing along the $z$-axis for clarity. This graph clearly shows that choosing non-sinusoidal matter deviations does not change the results significantly.}
\label{fig:cosh}
\end{figure}

After having analysed a variety of different sinusoidal single mode matter distributions and their coupling, we would like to investigate whether a more complex matter distribution would give a more significant deviation from the FLRW background. As shown in Fig.\ \ref{fig:dzda}, choosing a matter distribution which is not compensated, i.e.\ where integrating $F$ over any distance along the $z$-axis does not give zero, has significant effects on redshift and angular diameter distance. Now we would like to investigate whether compensated profiles which are more complex than simple single mode sinusoidal can have significant effects as well. To this end, we choose the initial profile for each over-density to take the form
\beq\label{eq:cosh}
\delta\propto \cosh^{-1}(\frac{z}{10{\text Mpc}})-C,
\eeq
where $C$ is a constant, and propagate light rays through a periodic array of such shapes. To ensure that this distribution is compensated, the constant $C$ needs chosen carefully. In Fig.\ \ref{fig:cosh} the results of this analysis are shown, where we have chosen the distance between the peaks to be $100$ Mpc today. The deviations from the FLRW results are not significantly different from the ones obtained by using sinusoidal distributions; in particular the dominant parameter in the effects on redshift and distances is the maximum length scale of the deviations.

\section{Conclusions}\label{sec:conc}
In this paper we have analysed the effects of non-linear structure on redshift and distance measures using the exact solution developed in our previous paper \citep{MeuBru11-1}. This model, described by the line element (\ref{eq:line}) in synchronous comoving coordinates, allows us to choose an arbitrary matter distribution along one line of sight with the growth rate of structure and the density distribution away from this axis being set by EFEs. A remarkable feature of our model is that the inhomogeneities are described by a single metric function $F$, which satisfies the same linear second order differential equation satisfied by the linear density perturbation $\delta$ in Newtonian perturbation theory as well as in relativistic perturbations when the synchronous comoving gauge is used. Therefore $F$ satisfies a superposition principle and, in particular, extends into the non-linear regime the same growing and decaying modes that $\delta$ shows in the linear regime.

We have developed the null geodesic equations and the Sachs optical equations in our model for light rays travelling in arbitrary directions. This set-up has then been applied to different physical situations, considering single sinusoidal mode deviations in the density, the coupling of two and three harmonic modes as well as a more complex matter distribution described by an array of peaks and voids along the line of sight. Furthermore, we have investigated which terms in the Sachs optical equations are mainly responsible for deviations from the FLRW values. Additionally, we have analysed the interaction of two and three modes in the growth of structure within our exact non-linear framework.

We consider the redshift and distance measures for single mode density deviations on different length scales and of different amplitudes. The largest effect for the redshift and distance measure to deviate from the FLRW results seems to be obtained for larger density deviations and larger scales. The results are shown in Figs.\ \ref{fig:dev500}, \ref{fig:dev100} and \ref{fig:dev1}. Given our conservative assumptions on the density profiles, all deviations are below the $1\%$ level. 

Even if the metric function $F$ satisfies a linear equation, our model is non-linear and so the effects of two modes on the redshift and distances does not need to be the same as their combined effects, hence we considered how modes interact in the growth of the density deviations and how the redshift and distance measures are affected by them. We find that combining small and large scale density deviations has significant effects on the growth of structure, in that the peaks of the large scale modes significantly enhance the growth of small scale deviations, see Fig.\ \ref{fig:deltas} for plots illustrating this point and the text in Sec.\ \ref{sec:res2}. For more details on the growth of structure in our model, see \citet{MeuBru11-1} and for a more general discussion of the relativistic behaviour of inhomogeneities and overall properties of the class of solutions we consider, see \citet{MatPanSae94-1,MatPanSae94,BruMatPan95-2,BruMatPan95-1}. The interaction between short and long wavelength modes observed in our model does not seem to have a significant effect on the redshift and distances measures though, as the effect of the long wavelength deviation remains dominant despite the presence of small scale deviations, see Fig.\ \ref{fig:dev1100} for the results of a $1$ Mpc and $100$ Mpc (today) mode combination. This implies that mode coupling does not provide significantly larger deviations on the redshift and distances than the individual modes.

To generalise our result to density distributions where peaks and voids are more pronounced than in a single mode sinusoidal, we have considered an array of density profiles given in Eq.\ (\ref{eq:cosh}), which provides quite peaked over-densities and large voids separating them, choosing a typical array scale of $100$ Mpc today. The results of the light tracing for this distribution is given in Fig.\ \ref{fig:cosh}, where the density distribution is shown in the top panel. Given this density profile, which is clearly non-sinusoidal for all times, the deviations from the FLRW redshift and distance measure are still comparable in amplitude to the results found in the single $100$ Mpc mode analysis, with the large scale deviations being below the $1\%$ level.

For all these different density distributions, we investigated which of the terms in the Sachs optical equations is dominant in providing the deviations, see Fig.\ \ref{fig:focus} for the different terms for a single mode deviation. We find that in all cases the Ricci focusing term is dominating, while the effect of the shear on the angular diameter distance seems to be vanishingly small. Due to its special geometric character (the space-time is Petrov type D), in our model the Weyl focusing is exactly zero along the $z$-axis of symmetry and it is sub-dominant with respect to the Ricci focusing in direction at an angle $\alpha$ with respect to the $z$-axis because of a $\sin^2(\alpha)$ factor. However, both the Ricci and Weyl focusing are proportional to the density deviation $\delta$ and so we may expect that in a more general space-time they would be of the same order.

All the above mentioned results are for density deviations which we refer to as compensated, i.e. where the metric function $F$ averages to zero along the line of sight at all times, so that the initially small density profile is also compensated. The results from considering density profiles which do not average to zero initially, however, are quite different. In Fig.\ \ref{fig:dzda}, we show that over-dense and under-dense lines of sight have significant effects on the redshift and angular diameter distance. This implies that if we, on average, observe along lines of sight which are more or less dense than the background, we may need to expect significant effects. Therefore it is important to understand whether the lines of sight we observe along are really average ``skewers" through the Universe matter distribution. It also emphasises the importance of identifying the correct background density with observations, as an over- or under-estimate may affect our interpretation of observational results significantly. Understanding these effects is crucial to our interpretation of cosmological observations and hence we leave a deeper analysis for future work.

Finally, part of the motivation for our work has come from the strong deviations from the standard FLRW results in \citet{CliFer09}, and therefore we would like to briefly compare our results here. Considering density distributions which are initially compensated along the line of sight cannot provide deviations in the redshift and distance measures as large as found in \citet{CliFer09}. This does seem to be in agreement with most Swiss-Cheese type analyses. However, considering lines of sight which have an average density which is lower than the background density can provide similar results as found in \citet{CliFer09}.

\section*{Acknowledgments}
We are supported by the STFC (grant no.\ ST/H002774/1). We would like to thank David Bacon, Krzysztof Bolejko, Timothy Clifton, Pedro Ferreira, Hubert Lampeitl, Roy Maartens and Guido Pettinari for useful discussions.

\appendix

\section{Tetrad transformations}\label{sec:trans}
Here we would like to present how to find the Weyl focusing term $\Psi_0$ for light rays travelling at an angle $\alpha$ from the $z$-axis. In the Newman-Penrose formalism, the five Weyl scalars ($\Psi_0$-$\Psi_4$) are contractions of the Weyl tensor with a complex null tetrad, $l^a$, $n^a$, $m^a$ and $\bar{m}^a$; for our notation see \citet{Cha92}. In our first paper, \citep{MeuBru11-1}, we have derived the Weyl scalars for the null tetrad 
\beqa
m^a&=&\frac{1}{\sqrt{2}}(0,\frac{1}{S},-i\frac{1}{S},0),\label{eq:tetm}\\
n^a&=&\frac{1}{\sqrt{2}}(1,0,0,-\frac{1}{SZ}),\label{eq:tetn}\\
l^a&=&\frac{1}{\sqrt{2}}(1,0,0,\frac{1}{SZ}),\label{eq:tetl}
\eeqa
where $S$ and $Z$ are the metric functions and $\bar{m}^a$ is simply the complex conjugate of $m^a$. Please note that we are using slightly different notation here than in the first paper, as we follow the exact notation of \citet{Cha92} here to avoid confusion when referring to this book. In this special null tetrad, the only non-zero Weyl scalar is $\Psi_2$. In this paper, we are interested in light tracing and we find that the tetrad vector $l^a=dx^a/d\lambda$ and so depending on the direction the light rays travel in, the vector $l^a$ is going to change. Hence we need to understand how changes in the complex null tetrad affect the Weyl scalars. Given the properties of complex null tetrads, there are only three distinct types of transformations, those of type I
\beqa
l^a\rightarrow l^a, &m^a&\rightarrow m^a+al^a, \bar{m}^a\rightarrow\bar{m}^a+a^*l^a,\nonumber\\
 {\text and} &n^a&\rightarrow n^a+a^*m^a+a\bar{m}^a+aa^*l^a,
\eeqa
of type II
\beqa
n^a \rightarrow n^a, &m^a& \rightarrow m^a+bn^a, \bar{m}^a\rightarrow \bar{m}^a+b^*n^a,\nonumber\\
 {\text and} &l^a&\rightarrow l^a+b^*m^a+b\bar{m}^a+bb^*n^a,
\eeqa
and of type III
\beqa
l^a \rightarrow A^{-1}l^a, &n^a&\rightarrow An^a, m^a\rightarrow e^{i\theta}m^a,\nonumber\\
 {\text and} &\bar{m}^a&\rightarrow e^{-i\theta}\bar{m}^a,
\eeqa
where $a$ and $b$ are complex and $A$ and $\theta$ are real valued functions. Each of these three rotations has the effect of mixing the Weyl scalars in a certain way, see \citet{Cha92} for the exact relations. To find the tetrad determined by Eqs.\ (\ref{eq:tet1}) and (\ref{eq:tet2}) from the above tetrad, we need to perform a combination of these transformations. We have used, in the given order, a transformation of type II with 
\beq
b=i\frac{\cos(\alpha)-1}{\sin(\alpha)},
\eeq
a transformation of type III with
\beq
A_1=2\frac{1-\cos(\alpha)}{\sin^2(\alpha)} \text{ and } \theta_1=0,
\eeq
a transformation of type I with
\beq
a=-i\frac{\cos(\alpha)-1}{\sin(\alpha)},
\eeq
and finally a transformation of type III with
\beq
A_2=\frac{1}{\sqrt{2}E} \text{ and } \theta_2=0.
\eeq
From these rotations, we obtain the null tetrad given in  Eqs.\ (\ref{eq:tet1}) and (\ref{eq:tet2}) from the null tetrad in Eqs.\ (\ref{eq:tetm}) -(\ref{eq:tetl}). Given the transformation rules of the Weyl scalars for the above rotations, we find all five Weyl scalars to be non-zero in general and specifically, we find that
\beq
\tilde{\Psi}_0 =-3\sin^2(\alpha)E^2\Psi_2,
\eeq
where the tilde denotes the quantity after the rotations. This is the Weyl focusing term for light rays travelling at an angle $\alpha$ to the $z$-axis.

\bibliography{ReferencesMaster}

\begin{thebibliography}{}

\bibitem[\protect\citeauthoryear{{Barausse}, {Matarrese} \&
  {Riotto}}{{Barausse} et~al.}{2005}]{BarMatRio05}
{Barausse} E.,  {Matarrese} S.,    {Riotto} A.,  2005, \prd, 71, 063537

\bibitem[\protect\citeauthoryear{{Barnes} \& {Rowlingson}}{{Barnes} \&
  {Rowlingson}}{1989}]{BarRow89}
{Barnes} A.,  {Rowlingson} R.~R.,  1989, Classical and Quantum Gravity, 6, 949

\bibitem[\protect\citeauthoryear{Barrow \& Stein-Schabes}{Barrow \&
  Stein-Schabes}{1984}]{BarSte84}
Barrow J.~D.,  Stein-Schabes J.,  1984, Phys. Lett., A103, 315

\bibitem[\protect\citeauthoryear{{Biswas} \& {Notari}}{{Biswas} \&
  {Notari}}{2008}]{BisNot08}
{Biswas} T.,  {Notari} A.,  2008, \jcap, 6, 21

\bibitem[\protect\citeauthoryear{{Bolejko} \& {C{\'e}l{\'e}rier}}{{Bolejko} \&
  {C{\'e}l{\'e}rier}}{2010}]{BolCel10}
{Bolejko} K.,  {C{\'e}l{\'e}rier} M.,  2010, \prd, 82, 103510

\bibitem[\protect\citeauthoryear{{Bolejko}, {Krasi{\'n}ski}, {Hellaby} \&
  {C{\'e}l{\'e}rier}}{{Bolejko} et~al.}{2009}]{BolKraHelCel09}
{Bolejko} K.,  {Krasi{\'n}ski} A.,  {Hellaby} C.,    {C{\'e}l{\'e}rier} M.,
  2009, {Structures in the Universe by Exact Methods: Formation, Evolution,
  Interactions}.
Cambridge University Press

\bibitem[\protect\citeauthoryear{{Bonvin}, {Durrer} \& {Gasparini}}{{Bonvin}
  et~al.}{2006}]{BonDurGas06}
{Bonvin} C.,  {Durrer} R.,    {Gasparini} M.~A.,  2006, \prd, 73, 023523

\bibitem[\protect\citeauthoryear{{Brouzakis}, {Tetradis} \&
  {Tzavara}}{{Brouzakis} et~al.}{2007}]{BroTetTza07}
{Brouzakis} N.,  {Tetradis} N.,    {Tzavara} E.,  2007, \jcap, 2, 13

\bibitem[\protect\citeauthoryear{{Brouzakis}, {Tetradis} \&
  {Tzavara}}{{Brouzakis} et~al.}{2008}]{BroTetTza08}
{Brouzakis} N.,  {Tetradis} N.,    {Tzavara} E.,  2008, \jcap, 4, 8

\bibitem[\protect\citeauthoryear{{Bruni}, {Matarrese} \& {Pantano}}{{Bruni}
  et~al.}{1995a}]{BruMatPan95-2}
{Bruni} M.,  {Matarrese} S.,    {Pantano} O.,  1995a, Phys. Rev Let., 74, 1916

\bibitem[\protect\citeauthoryear{{Bruni}, {Matarrese} \& {Pantano}}{{Bruni}
  et~al.}{1995b}]{BruMatPan95-1}
{Bruni} M.,  {Matarrese} S.,    {Pantano} O.,  1995b, Astrophys. J., 445, 958

\bibitem[\protect\citeauthoryear{{Buchert}}{{Buchert}}{2008}]{Buc08}
{Buchert} T.,  2008, General Relativity and Gravitation, 40, 467

\bibitem[\protect\citeauthoryear{{Chandrasekhar}}{{Chandrasekhar}}{1992}]{Cha9%
2}
{Chandrasekhar} S.,  1992, {The mathematical theory of black holes}.
Oxford University Press

\bibitem[\protect\citeauthoryear{{Clifton}}{{Clifton}}{2011}]{Cli10}
{Clifton} T.,  2011, Classical and Quantum Gravity, 28, 164011

\bibitem[\protect\citeauthoryear{{Clifton} \& {Ferreira}}{{Clifton} \&
  {Ferreira}}{2009a}]{CliFer09}
{Clifton} T.,  {Ferreira} P.~G.,  2009a, \prd, 80, 103503

\bibitem[\protect\citeauthoryear{{Clifton} \& {Ferreira}}{{Clifton} \&
  {Ferreira}}{2009b}]{Clifer09-2}
{Clifton} T.,  {Ferreira} P.~G.,  2009b, \jcap, 10, 26

\bibitem[\protect\citeauthoryear{{Clifton} \& {Zuntz}}{{Clifton} \&
  {Zuntz}}{2009}]{CliZun09}
{Clifton} T.,  {Zuntz} J.,  2009, \mnras, 400, 2185

\bibitem[\protect\citeauthoryear{{Dyer} \& {Roeder}}{{Dyer} \&
  {Roeder}}{1972}]{DyeRoe72}
{Dyer} C.~C.,  {Roeder} R.~C.,  1972, \apjl, 174, L115+

\bibitem[\protect\citeauthoryear{{Dyer} \& {Roeder}}{{Dyer} \&
  {Roeder}}{1973}]{DyeRoe73}
{Dyer} C.~C.,  {Roeder} R.~C.,  1973, \apjl, 180, L31+

\bibitem[\protect\citeauthoryear{{Dyer} \& {Roeder}}{{Dyer} \&
  {Roeder}}{1974}]{DyeRoe74}
{Dyer} C.~C.,  {Roeder} R.~C.,  1974, \apj, 189, 167

\bibitem[\protect\citeauthoryear{{Etherington}}{{Etherington}}{1933}]{Eth33}
{Etherington} I.~M.~H.,  1933, Philosophical Magazine, 15, 761

\bibitem[\protect\citeauthoryear{{Futamase} \& {Sasaki}}{{Futamase} \&
  {Sasaki}}{1989}]{FutSas89}
{Futamase} T.,  {Sasaki} M.,  1989, \prd, 40, 2502

\bibitem[\protect\citeauthoryear{{Goode} \& {Wainwright}}{{Goode} \&
  {Wainwright}}{1982a}]{GooWai82}
{Goode} S.~W.,  {Wainwright} J.,  1982a, \mnras., 198, 83

\bibitem[\protect\citeauthoryear{{Goode} \& {Wainwright}}{{Goode} \&
  {Wainwright}}{1982b}]{GooWai82-1}
{Goode} S.~W.,  {Wainwright} J.,  1982b, Phys. Rev. D, 26, 3315

\bibitem[\protect\citeauthoryear{{Krasinski}}{{Krasinski}}{1997}]{Kra97}
{Krasinski} A.,  1997, {Inhomogeneous Cosmological Models}.
Cambridge University Press

\bibitem[\protect\citeauthoryear{{Lindquist} \& {Wheeler}}{{Lindquist} \&
  {Wheeler}}{1957}]{LinWhe57}
{Lindquist} R.~W.,  {Wheeler} J.~A.,  1957, Rev. Mod. Phys., 29, 432

\bibitem[\protect\citeauthoryear{{Marra}, {Kolb}, {Matarrese} \&
  {Riotto}}{{Marra} et~al.}{2007}]{Mar07}
{Marra} V.,  {Kolb} E.~W.,  {Matarrese} S.,    {Riotto} A.,  2007, \prd, 76,
  123004

\bibitem[\protect\citeauthoryear{{Matarrese}, {Pantano} \& {Saez}}{{Matarrese}
  et~al.}{1994a}]{MatPanSae94-1}
{Matarrese} S.,  {Pantano} O.,    {Saez} D.,  1994a, \mnras., 271, 513

\bibitem[\protect\citeauthoryear{{Matarrese}, {Pantano} \& {Saez}}{{Matarrese}
  et~al.}{1994b}]{MatPanSae94}
{Matarrese} S.,  {Pantano} O.,    {Saez} D.,  1994b, Phys. Rev. Let., 72, 320

\bibitem[\protect\citeauthoryear{{Meures} \& {Bruni}}{{Meures} \&
  {Bruni}}{2011}]{MeuBru11-1}
{Meures} N.,  {Bruni} M.,  2011, \prd, 83, 123519

\bibitem[\protect\citeauthoryear{{Nwankwo}, {Ishak} \& {Thompson}}{{Nwankwo}
  et~al.}{2011}]{NwaIshTho11}
{Nwankwo} A.,  {Ishak} M.,    {Thompson} J.,  2011, \jcap, 5, 28

\bibitem[\protect\citeauthoryear{{Peebles}}{{Peebles}}{1980}]{Pee80}
{Peebles} P.~J.~E.,  1980, {The large-scale structure of the universe}.
Princeton University Press

\bibitem[\protect\citeauthoryear{{Pyne} \& {Birkinshaw}}{{Pyne} \&
  {Birkinshaw}}{2004}]{PynBir04}
{Pyne} T.,  {Birkinshaw} M.,  2004, \mnras, 348, 581

\bibitem[\protect\citeauthoryear{{R{\"a}s{\"a}nen}}{{R{\"a}s{\"a}nen}}{2006}]{%
Ras06}
{R{\"a}s{\"a}nen} S.,  2006, JCAP, 11, 3

\bibitem[\protect\citeauthoryear{R{\"a}s{\"a}nen}{R{\"a}s{\"a}nen}{2011}]{Ras1%
1-2}
R{\"a}s{\"a}nen S.,  2011, arXiv:astro-ph.CO/1107.1176

\bibitem[\protect\citeauthoryear{{Redmount}}{{Redmount}}{1988}]{Red88}
{Redmount} I.~H.,  1988, \mnras, 235, 1301

\bibitem[\protect\citeauthoryear{{Sachs}}{{Sachs}}{1961}]{Sac61}
{Sachs} R.,  1961, Royal Society of London Proceedings Series A, 264, 309

\bibitem[\protect\citeauthoryear{{Sarkar}, {Yadav}, {Pandey} \&
  {Bharadwaj}}{{Sarkar} et~al.}{2009}]{Sar09}
{Sarkar} P.,  {Yadav} J.,  {Pandey} B.,    {Bharadwaj} S.,  2009, \mnras., 399,
  L128

\bibitem[\protect\citeauthoryear{{Sasaki}}{{Sasaki}}{1987}]{Sas87}
{Sasaki} M.,  1987, \mnras, 228, 653

\bibitem[\protect\citeauthoryear{Springel et~al.,}{Springel
  et~al.}{2005}]{Spr05}
Springel V.,  et~al., 2005, \nat, 435, 629

\bibitem[\protect\citeauthoryear{{Stephani}, {Kramer}, {MacCallum},
  {Hoenselaers} \& {Herlt}}{{Stephani} et~al.}{2003}]{Ste03}
{Stephani} H.,  {Kramer} D.,  {MacCallum} M.,  {Hoenselaers} C.,    {Herlt} E.,
   2003, {Exact solutions of Einstein's field equations}.
Cambridge University Press

\bibitem[\protect\citeauthoryear{{Szekeres}}{{Szekeres}}{1975}]{Sze75}
{Szekeres} P.,  1975, Com. in Math. Phys., 41, 55

\bibitem[\protect\citeauthoryear{{Szybka}}{{Szybka}}{2011}]{Szy11}
{Szybka} S.~J.,  2011, \prd, 84, 044011

\end{thebibliography}
\bibliographystyle{mn2e}
\label{lastpage}
\end{document}